\documentclass[12pt, oneside, onecolumn, reqno, notitlepage, centertags]{amsart}
\usepackage[tmargin=32mm,bmargin=32mm,lmargin=32mm,rmargin=32mm]{geometry}
\usepackage[english]{babel}
\usepackage[utf8]{inputenc}
\usepackage{comment}
\usepackage{rotating} 
\usepackage{booktabs}
\usepackage{etoolbox}
\usepackage{multirow}
\usepackage{setspace} 
\usepackage{float} 
\usepackage{subcaption} 
\usepackage{bm} 
\usepackage{mathtools} 
\usepackage{xcolor}
\usepackage{epstopdf}
\usepackage{longtable}
\usepackage{tabularx}
\usepackage{threeparttable}
\usepackage{microtype}
\usepackage{tgtermes} 
\usepackage{siunitx}

\usepackage{caption}
\DeclareMathAlphabet{\mathpzc}{OT1}{pzc}{m}{it}

\usepackage{dsfont}
\newcommand{\E}{\mathds{E}} 
 
\newcommand{\re}{\mathds{R}}

\DeclareMathOperator{\diag}{diag}

\DeclareMathOperator{\Var}{Var}

\usepackage{amssymb,amsthm,amsfonts,amsmath}
\usepackage{enumerate}

\usepackage{rotate,graphicx,hyperref}
\usepackage{tgtermes} 
\usepackage{color} 
\usepackage{epstopdf}
\usepackage[round]{natbib}
\usepackage{booktabs} 
\usepackage{tablefootnote}

\allowdisplaybreaks[4]

\begin{document}

\title[A Beta-Based Heteroskedasticity-Consistent Covariance Matrix Estimator]{A Beta-Based Heteroskedasticity-Consistent Covariance Matrix Estimator} 

\author{Marina O.~Cunha}

\address{Departamento de Estat\'{\i}stica,
		Universidade Federal de Pernambuco\\
		Recife/PE, 58051-900, Brazil\\
        ORCID: 0009-0007-4938-3881\\
        E-mail: marina.oliveirac@ufpe.br}

\author{Francisco Cribari-Neto}

\address{Departamento de Estat\'{\i}stica,
		Universidade Federal de Pernambuco\\
		Recife/PE, 58051-900, Brazil\\
        ORCID: 0000-0002-5909-6698\\
		E-mail: francisco.cribari@ufpe.br\\
		(Corresponding author)}

\author{Pedro R.~D.~Marinho}
\address{Departamento de Estat\'{\i}stica,
		Universidade Federal da Para\'{\i}ba\\
		Jo\~{a}o Pessoa/PB, 58051-900, Brazil\\
        ORCID: 0000-0003-1591-8300\\
        E-mail: pedro.rafael.marinho@gmail.com}


\maketitle

\begin{abstract}
This paper introduces an adaptive leverage correction for heteroskedasticity-consistent covariance matrix estimation in ordinary least squares regression. Unlike existing estimators, which rely on predetermined adjustment functions, the proposed approach uses a fitted Beta distribution calibrated to the design matrix's empirical leverage structure. Its parameters are estimated from observed leverage values, allowing the adjustment to reflect the overall configuration rather than individual values alone. The estimator accommodates heterogeneous leverage patterns while avoiding excessive growth of adjustment factors. Monte Carlo simulations show accurate finite-sample inference and confidence interval coverage, alongside desirable asymptotic properties. Empirical applications illustrate its advantages with influential observations and overshooting by existing estimators. The open-source \textsf{R} package \textsl{hcinfer} implements the estimator and related procedures.
\end{abstract}

\section{Introduction}\label{S:introduction} 

Linear regression models are among the most widely used statistical tools for analyzing relationships between a continuous response variable and one or more explanatory variables. Parameter estimation is typically carried out by ordinary least squares (OLS), a method that does not require distributional assumptions on the regression errors. For statistical inference --- including hypothesis testing and the construction of confidence intervals --- it is common to assume that the errors are homoskedastic and normally distributed. While normality is often invoked primarily for convenience, the assumption of homoskedasticity is far more consequential and is frequently violated in cross-sectional applications.

When heteroskedasticity is present, the OLS estimator of the regression coefficients remains unbiased, consistent, and asymptotically normal under standard regularity conditions. However, the conventional estimator of its covariance matrix becomes biased and inconsistent, invalidating standard inferential procedures. As a result, statistical inference based on classical standard errors can be severely misleading in the presence of heteroskedasticity of unknown form.

A large literature has therefore developed methods for estimation and inference in linear regression models under heteroskedasticity; see, among others, \citet{Cribari+Pereira_2019} and the recent comprehensive review by \citet{Farrar+et-al_2025}. One particularly influential and practically appealing approach relies on heteroskedasticity-consistent (HC) estimators of the covariance matrix of the OLS estimator. These estimators yield asymptotically valid standard errors that can be used to conduct quasi-$t$ (or quasi-$z$) hypothesis tests on the regression coefficients. Because they do not require explicit modeling of the variance function, HC estimators have become a standard tool in empirical work, especially in settings where the form of heteroskedasticity is unknown or difficult to specify \citep{Hayes+Cai_2007}.

Over the past decades, several HC covariance matrix estimators have been proposed, differing mainly in how they adjust squared OLS residuals to account for leverage. Such adjustments are particularly important in samples of moderate size, where high-leverage observations may exert a disproportionate influence on inference. Most existing HC estimators employ relatively simple functional forms for these leverage adjustments, often involving powers of $(1-h_t)$, where $h_t$ denotes the leverage of the $t$th observation. While more aggressive adjustments are sometimes effective at improving Type~I error control, they can also lead to extremely large correction factors for high-leverage observations. This behavior may lead to excessively inflated standard errors and unstable inference. In this paper, we identify and label this phenomenon as \emph{overshooting}: a previously underappreciated effect whereby aggressive leverage corrections induce disproportionate inflation of residual-based adjustments under strong leverage.

Recent simulation-based evidence reinforces these concerns. As documented by \citet{Rajh+et-al_2025}, the finite-sample performance of widely used HC estimators --- as well as related bootstrap procedures --- can be highly sensitive to sample size, the degree of heteroskedasticity, and the presence of leverage points. In particular, strong leverage combined with small or moderate sample sizes may induce erratic behavior in standard errors and hypothesis tests, compromising both size and power properties. These findings highlight the need for heteroskedasticity-consistent covariance matrix estimators that balance effective leverage correction with numerical stability.

Against this background, the present paper introduces a new heteroskedasticity-consistent covariance matrix estimator for OLS regression, denoted $\text{HC}_\beta$. Like existing HC estimators, the proposed method incorporates a leverage-based adjustment to squared residuals. Its key novelty lies in how this adjustment is constructed. Rather than relying solely on fixed or piecewise-defined exponents of $(1-h_t)$, $\text{HC}_\beta$ employs an adjustment factor derived from a Beta distribution whose parameters are estimated from the empirical distribution of the leverage values. This data-driven mechanism allows the adjustment to flexibly adapt to the overall leverage structure of the design matrix, accommodating asymmetry and heterogeneity across observations while avoiding explosive behavior under  extreme leverage.

By construction, the proposed estimator moderates the correction applied to high-leverage observations without reverting to overly simplistic adjustments. As a result, $\text{HC}_\beta$ delivers stable and reliable standard errors in finite samples, correcting the well-known downward bias of basic HC estimators while preventing the overshooting often observed in more aggressive alternatives. This balance leads to more coherent and robust inference on regression coefficients, even in empirical settings characterized by strong leverage and unknown heteroskedasticity.

The use of the Beta distribution is not essential to the proposed framework. In principle, other parametric families defined on the unit interval could also be employed to construct leverage-based adjustment factors. Examples include the Kumaraswamy distribution \citep{Jones_2009}, whose analytical tractability and simple closed-form distribution function make it an attractive alternative. The $\text{Beta}(a,b)$ distribution was chosen because it combines several desirable features in the present context. First, it is indexed by two shape parameters, providing substantially greater flexibility than one-parameter alternatives. Second, its parameters can be estimated directly from the leverage values by simple closed-form method-of-moments estimators, avoiding additional numerical optimization. Third, the uniform distribution, which corresponds to the special case $a=b=1$, is nested within the Beta family, allowing the proposed approach to accommodate designs exhibiting approximately uniform leverage patterns as well as highly asymmetric configurations. Finally, the Beta family is invariant under the transformation $z \mapsto 1-z$: if $Z \sim \text{Beta}(a,b)$, then $1-Z \sim \text{Beta}(b,a)$. Consequently, working with the complementary leverage values $u_t = 1-h_t$ rather than with the leverage values $h_t$ simply interchanges the two shape parameters, preserving the same modeling framework and interpretation of the adjustment function. These features make the Beta family a natural starting point for developing adaptive heteroskedasticity-consistent corrections.

Throughout the paper, we follow the conventional fixed-design regression framework adopted, for example, by \citet{MacKinnon+White_1985}, \citet{Cribari+Ferrari+Cordeiro_2000}, and \citet{Vasconcellos+Cribari_2026}, in which the design matrix $X$ is treated as fixed. Consequently, the leverage values $h_t$ and the complementary leverage values $u_t = 1 - h_t$ are deterministic functions of the observed design matrix rather than realizations of random variables. The fitted Beta distribution introduced below is therefore not intended as a probabilistic model for the latter. Instead, it serves as a flexible parametric device for calibrating the adjustment factor to the empirical leverage configuration of the observed design matrix.

A key distinction between the proposed approach and existing HC estimators is that the correction is calibrated from the entire empirical distribution of the leverage values, rather than from each $h_t$ in isolation, as in HC2--HC5m, whose differences reduce to the choice of function and tuning constants applied pointwise to $(1-h_t)$. This shift from a pointwise to a distribution-based calibration constitutes the main conceptual contribution of the paper, and provides a practical mechanism for moderating overshooting (documented in Sections~\ref{S:simulation} and~\ref{S:applications}) while preserving the asymptotic properties of heteroskedasticity-consistent inference.

An \textsf{R} package, \textsl{hcinfer}, was developed to accompany the methodology proposed in this article. The package implements the $\text{HC}_\beta$ estimator as well as several classical heteroskedasticity-consistent covariance matrix estimators, providing a unified framework for inference in linear regression models. It includes routines for covariance matrix estimation, hypothesis testing, confidence interval construction, diagnostic visualization, and comparative analysis of adjustment factors associated with leverage values.

The remainder of the paper is organized as follows. Section~\ref{S:model and estimators} presents the regression framework and reviews standard heteroskedasticity-consistent covariance matrix estimators. Section~\ref{S:hc-beta} introduces the proposed $\text{HC}_\beta$ estimator and discusses its main properties. Section~\ref{S:simulation} reports Monte Carlo simulation results assessing its finite-sample performance relative to existing methods. Section~\ref{S:applications} provides empirical applications illustrating the practical advantages of the new estimator. The implementation of the proposed estimator in the \textsf{R} statistical computing environment is briefly described in Section~\ref{S:r-package}. Section~\ref{S:conclusions} concludes and outlines directions for future research.

\section{Model and estimators}\label{S:model and estimators} 

Consider the linear regression model
\begin{equation*}
\bm{y} = X\bm{\beta} + \bm{e},
\end{equation*}
where $\bm{y} = (y_1, \ldots, y_n)'$ is an $n \times 1$ vector of observations on the dependent variable, $X$ is an $n \times p$ matrix of regressors with full column rank ($\mathrm{rank}(X) = p < n$), $\bm{\beta}=(\beta_1, \ldots, \beta_p)'$ is a $p \times 1$ vector of unknown parameters, and $\bm{e}=(e_1, \ldots, e_n)'$ is an $n \times 1$ vector of uncorrelated errors. The errors have mean zero and finite variance, that is, $\E(e_t) = 0$ and $\Var(e_t) = \sigma_t^2 \in (0, \infty)$ for $t = 1, \ldots, n$. The covariance matrix of $\bm{e}$ is $\Omega = \diag\{ \sigma_t^2\}$. When the errors are homoskedastic ($\sigma_t^2 = \sigma^2$ for all $t$), we have $\Omega = \sigma^2 I_n$, where $I_n$ is the identity matrix of order $n$.

Let $\bm{x}_t$ denote the $t$th row of $X$ expressed as a column vector. The leverage of observation $t$ is $h_t = \bm{x}_t' (X'X)^{-1} \bm{x}_t$. It quantifies the influence of the design point on its own fitted value and, equivalently, measures how far the corresponding observation lies from the center of the predictor space. It satisfies $0 \le h_t \le 1$, has average value $\bar h = p/n$, and is commonly used to identify high-leverage observations. Throughout the paper, an observation is classified as a leverage point whenever $h_t > 3\bar h$, a widely used rule of thumb. We also define $u_t = 1 - h_t$, which we refer to as the \emph{complementary leverage} of observation $t$.

The ordinary least squares estimator of $\bm{\beta}$ is $\bm{\hat{\beta}} = (X'X)^{-1}X'\bm{y}$, which is unbiased, that is, $\E(\bm{\hat{\beta}}) = \bm{\beta} \; \forall \bm{\beta} \in \re^p$. Its covariance matrix is given by
\begin{equation*}
\Psi = (X'X)^{-1} X' \Omega X (X'X)^{-1}.
\end{equation*} 
Under homoskedasticity, we have $\Psi = \sigma^2 (X'X)^{-1}$, which can be conveniently estimated by $\hat{\sigma}^2 (X'X)^{-1}$, where $\hat{\sigma}^2 = \hat{\bm{e}}'\hat{\bm{e}} / (n - p)$. Here, $\hat{\bm{e}} = \bm{y} - X\bm{\hat{\beta}}$ denotes the vector of ordinary least squares residuals. It is, however, desirable to estimate $\Psi$ consistently both under homoskedasticity and under heteroskedasticity of unknown form. To address this challenge, a broad class of heteroskedasticity-consistent estimators has been proposed in the literature, providing consistent estimation of $\Psi$ without requiring specific assumptions about the structure of heteroskedasticity. These estimators share the general structure
\begin{equation*}
\widehat{\Psi}_{\text{HC}} = (X'X)^{-1} X' \widehat{\Omega} X (X'X)^{-1},
\end{equation*}
where
\begin{equation*}
\widehat{\Omega} = \diag\{\hat{e}_t^2 \, g_t\},
\end{equation*}
and $\hat{e}_t = y_t - \bm{x}_t'\bm{\hat{\beta}}$ is the $t$th OLS residual,
with $g_t$ denoting the estimator-specific adjustment factor, which satisfies $g_t \to 1$ as $n \to \infty$ for each fixed $t$.

The various HC estimators differ primarily in their choice of the adjustment factor $g_t$, which governs the extent to which the residuals are scaled according to their leverage values. The simplest form, HC0 \citep{White_1980}, sets $g_t = 1$ for all $t$, ignoring leverage entirely. HC1 \citep{Hinkley_1977} applies a uniform scaling, $g_t = n/(n-p)$. HC2 \citep{MacKinnon+White_1985} and HC3 \citep{Davidson+MacKinnon_1993} introduce leverage-based corrections, using $g_t = 1/(1-h_t)$ and $g_t = 1/(1-h_t)^2$, respectively. More sophisticated adjustments were later proposed: HC4 \citep{Cribari_2004} replaces the fixed exponent with an adaptive one, $g_t = (1-h_t)^{-\delta_t}$, where $\delta_t = \min\{4, h_t/\bar{h}\}$; and HC4m \citep{Cribari+Silva_2011} refines this idea by setting $\delta_t = \min\{1, h_t/\bar{h}\} + \min\{1.5, h_t/\bar{h}\}$.

HC0 is known to exhibit a downward bias --- it often underestimates the true variances, particularly in small samples or when leverage points are present. As pointed out by \citet{Chesher+Jewitt_1987}, this bias can be substantial when some $h_t$ values are large. In such cases, the corresponding least squares residuals tend to be small on average, and since the HC0 estimator interprets small residuals as indicative of low error variances, it systematically understates the true variability.

HC estimators are widely used to conduct inference in linear regression via quasi-$t$ (or quasi-$z$) hypothesis tests. Consider testing the null hypothesis $\mathcal{H}_0\colon \beta_j = \beta_{j}^{(0)}$ against the two-sided alternative $\mathcal{H}_1\colon \beta_j \not= \beta_{j}^{(0)}$, for $j = 1, \ldots, p$. The corresponding test statistic is
\begin{equation*}
t_j = \frac{\hat{\beta}_j - \beta_{j}^{(0)}}{\sqrt{\widehat{\Psi}_{\text{HC}_{jj}}}},
\end{equation*}
where $\widehat{\Psi}_{\text{HC}_{jj}}$ denotes the $j$th diagonal element of a heteroskedasticity-consistent covariance matrix estimator. Under $\mathcal{H}_0$, $t_j$ is asymptotically standard normal. The null hypothesis is rejected at significance level $\alpha \in (0,1)$ whenever $|t_j| > z_{1-\alpha/2}$, where $z_{1-\alpha/2}$ denotes the $1-\alpha/2$ upper quantile of the standard normal distribution.

Many HC estimators have been proposed in the literature, including some based on non-ordinary least squares residuals \citep{Furno_1996b} and others that incorporate bias adjustments \citep{Cribari+Ferrari+Cordeiro_2000}. For a comprehensive review of this literature, we refer the reader to \cite{Farrar+et-al_2025}. To maintain focus in the present work, we restrict our attention to the estimators introduced above, namely HC0, HC1, HC2, HC3, HC4, and HC4m. 

\section{The $\text{HC}_\beta$ estimator}\label{S:hc-beta} 

This section presents a novel approach to heteroskedasticity-consistent covariance matrix estimation. We first develop the proposed $\text{HC}_\beta$ estimator and discuss its main features. We then establish its consistency, showing that it is first-order asymptotically equivalent to conventional heteroskedasticity-consistent covariance matrix estimators under standard regularity conditions.

\subsection{The $\text{HC}_\beta$ estimator}

A key observation is that the term $1-h_t$ in existing HC estimators can be interpreted as the cumulative distribution function (CDF) of a Uniform$(0,1)$ distribution evaluated at $1-h_t$, with $1-h_t \in [0,1]$. That is,
\begin{equation*}
F_{\text{Uniform}}(1-h_t) = 1 - h_t, \quad 0 \le 1-h_t \le 1.
\end{equation*}
This observation motivates replacing the uniform CDF with the CDF of a more flexible $\text{Beta}(a,b)$ distribution, of which the Uniform$(0,1)$ distribution is a special case (obtained when $a=b=1$),
\begin{equation*}
F_{\text{Beta}}(1-h_t; a, b) = I_{1-h_t}(a,b),
\end{equation*}
where $I_v(a,b)$ denotes the regularized incomplete Beta function (also called the normalized incomplete Beta function), defined as
\begin{equation*}
I_v(a,b) = \int_0^v \frac{t^{a-1}(1-t)^{b-1}}{B(a,b)} \, dt, \quad 0 \le v \le 1,
\end{equation*}
$B(a,b) = [\Gamma(a)\Gamma(b)]/\Gamma(a+b)$ being the Beta function and $\Gamma(\cdot)$ being the gamma function. The shape parameters $a$ and $b$ are estimated from the empirical distribution of the leverage values by the method of moments. Because the Beta family spans a broad range of shapes, its cumulative distribution function provides a flexible data-adaptive transformation of the leverage values. The key innovation of $\text{HC}_\beta$ is twofold: (i)~the adjustment factor is based on a moment-matched Beta distribution rather than a fixed functional form and (ii)~the Beta parameters are estimated adaptively from the leverage values, allowing the adjustment scheme to adjust to the specific structure of the design matrix.

The proposed adjustment does not rely on any probabilistic model for the
leverage values. Instead, the Beta family is introduced solely as a flexible
parametric framework for constructing the adjustment factor. The shape
parameters $a$ and $b$ are obtained by matching the first two sample moments of
the complementary leverage values $u_t = 1-h_t$. Consequently, the resulting
method-of-moments estimators should not be interpreted as estimates of
population parameters, nor is it assumed that the complementary leverage values
follow a Beta distribution. Rather, they identify the member of the Beta family
whose first two moments coincide with those of the observed leverage
configuration, thereby calibrating the adjustment function to the geometry of
the design matrix rather than modeling the leverage distribution itself.

Leverage values $h_t$ are theoretically bounded in $[0,1]$ and can reach or approach $1$ very closely, particularly for high-leverage observations. Consequently, $u_t = 1 - h_t$ can be exactly zero or approach $0$, producing boundary values --- points at or extremely close to the lower limit of the unit interval --- where the Beta CDF changes rapidly. When the Beta CDF is evaluated at such extreme arguments, it takes values very close to zero (or exactly zero), which would make its reciprocal (later used as part of the adjustment applied to the $t$th squared residual) extremely large or undefined. Such pathological adjustments could unduly amplify the heteroskedasticity correction and distort inference.

To prevent this, we truncate $u_t$ below at $0.01$ and above at $0.99$, defining
\[
w_t=\max\{0.01,\min\{u_t,0.99\}\}.
\]
Using $w_t$ instead of $u_t$ excludes the extreme boundary regions of the unit
interval, thereby avoiding excessively large adjustment factors and improving
numerical stability.

The truncation also regularizes the estimation of the Beta shape parameters. As $n\rightarrow\infty$ with $p$ fixed, the leverage values satisfy $h_t\rightarrow0$ for every $t$, implying that $u_t\rightarrow1$ and that the empirical variance of the complementary leverage values becomes increasingly small. Since the Beta shape parameters are inversely related to this variance, moment matching would require increasingly large values of both $a$ and $b$, producing a highly concentrated Beta distribution near the upper endpoint of the unit interval. Evaluating the Beta cumulative distribution function in this regime may lead to numerical instability and unnecessarily erratic finite-sample adjustments. By maintaining a small but positive distance from the boundaries, the proposed truncation avoids this degeneracy while preserving the adaptive nature of the correction.


The proposed procedure is not particularly sensitive to the specific choice of truncation limits. As a robustness check, we repeated the Monte Carlo experiments presented in Section~\ref{S:simulation} and the empirical analyses in Section~\ref{S:applications} using the alternative intervals $(0.001,0.999)$ and $(0.02,0.98)$. The results (not reported for brevity) were virtually indistinguishable from those obtained with the default truncation limits: empirical rejection rates, confidence interval coverages, and the substantive conclusions remained essentially unchanged. This indicates that the role of truncation is simply to prevent pathological evaluations of the Beta CDF near the boundaries of the unit interval, rather than to determine the inferential behavior of the proposed estimator.

The parameters $a$ and $b$ are estimated by the method of moments, which has
the advantages of a closed-form solution and does not require numerical
optimization. The resulting estimates should not be interpreted as estimators
of population parameters; rather, they determine the member of the Beta family
whose first two moments match those of the observed leverage configuration.
The moment estimators are given by
\begin{equation*}
\hat{a} = \hat{\mu}_\beta \, \hat{\phi}_\beta
\,\, \text{ and } \,\,
\hat{b} = (1 - \hat{\mu}_\beta) \, \hat{\phi}_\beta,
\end{equation*}
where $\hat{\mu}_\beta = n^{-1} \sum_{t=1}^n w_t$ and
$\hat{\phi}_\beta = \hat{\mu}_\beta(1 - \hat{\mu}_\beta)/s_w^2 - 1$, with
$s_w^2 = (n-1)^{-1} \sum_{t=1}^n (w_t - \hat{\mu}_\beta)^2$.
To ensure that the estimators are well-defined for all possible empirical leverage configurations, we adopt the following convention: if the empirical variance $s_w^2$ is exactly zero (which occurs asymptotically when the maximum leverage converges to zero and all truncated values $w_t$ become equal to $0.99$), we define $\hat{a} = \hat{b} = +\infty$. Under this degenerate case, the regularization mechanism described below guarantees that the final shape parameters $\tilde{a}$ and $\tilde{b}$ are bounded and strictly well-defined.

To improve numerical stability in small samples, we apply a
simple shrinkage regularization toward the Uniform$(0,1)$ distribution, which
corresponds to the Beta distribution with $a=b=1$. The uniform distribution is
used as the shrinkage target because it represents the baseline case in which
the proposed adjustment reduces to the classical HC0 correction. Thus, the
regularization acts only as a safeguard against unstable moment estimates in
small samples, while preserving the interpretation of the proposed estimator
as an adaptive extension of HC0. We use
\begin{align*}
\zeta &= \frac{n}{n + 50}, \\
\tilde{a} &= \min\Bigl\{ \max\bigl\{ (1 - \zeta) + \zeta \hat{a}, \, \epsilon \bigr\}, \, A_{\max} \Bigr\}, \\
\tilde{b} &= \min\Bigl\{ \max\bigl\{ (1 - \zeta) + \zeta \hat{b}, \, \epsilon \bigr\}, \, B_{\max} \Bigr\},
\end{align*}
where $\epsilon > 0$ is a small constant that ensures the shape parameters remain strictly positive, preventing the Beta distribution from becoming undefined. In our implementation, we set $\epsilon = 0.01$, $A_{\max} = 10{,}000$, and $B_{\max} = 10{,}000$.

The lower bound $\epsilon$ preserves the data-adaptive nature of the method by allowing the parameters to reflect highly skewed empirical leverage configurations (where moment estimates may naturally fall below unity).
The upper bounds, $A_{\max}$ and $B_{\max}$, serve a dual theoretical and
numerical purpose. Asymptotically, as $n \to \infty$ with $p$ fixed, the
empirical variance of the transformed leverages vanishes, which would
otherwise cause the moment estimates to diverge to infinity. The upper bounds
ensure that the parameter space remains compact, a crucial requirement for
the asymptotic consistency argument developed below. In finite samples, they
prevent the parameters from becoming excessively large, thereby avoiding
numerical underflow or overflow when evaluating the Beta cumulative
distribution function near the boundaries of the unit interval. By selecting
a sufficiently large upper bound, we preserve the data-adaptive nature of the
method and avoid the artificial distortion of the empirical leverage structure
that would result from overly restrictive caps.

The shrinkage weight $\zeta = n/(n+50)$ is an empirical regularization rule rather than a quantity derived from asymptotic theory. It was selected after extensive Monte Carlo experiments as providing a favorable compromise between stability and adaptivity across a wide range of sample sizes and leverage configurations. When $n=50$, equal weight is assigned to the empirical moment estimates and the uniform target, whereas the influence of the target decreases smoothly as the sample size increases. Consequently, the adjustment becomes progressively more data-driven while retaining additional stability in small samples. Although other shrinkage functions could be employed, our simulation results indicate that this simple specification performs well over the scenarios considered in this paper.

The uniform distribution ($a=b=1$) serves as a natural reference point because it yields $F_{\text{Beta}}(w_t; 1, 1) = w_t$. This ensures that the shrinkage target corresponds exactly to the baseline leverage correction used in conventional heteroskedasticity-consistent estimators, such as HC2, HC3, HC4, and HC4m. Consequently, the regularization not only stabilizes the parameter estimates in finite samples but also guarantees that the proposed method remains a coherent, data-adaptive extension of the classical HC framework.

The proposed adjustment follows the same principle underlying the family of HC estimators, namely, correcting the squared residuals by a monotone function of the leverage values to account for the leverage-induced downward bias of the OLS residual variance. Existing HC estimators differ primarily in the choice of this correction function. Whereas HC2, HC3, HC4, and related estimators employ predetermined algebraic transformations of $(1-h_t)$, the proposed estimator replaces these fixed forms with a smooth data-adaptive transformation based on the cumulative distribution function of a fitted Beta distribution. This construction preserves the monotonicity required of a leverage correction while allowing its growth to adapt to the empirical leverage configuration.

The $\text{HC}_\beta$ covariance matrix estimator is defined as
\begin{equation*}
\widehat{\Psi}_{\text{HC}_\beta} = (X'X)^{-1}X' \widehat{\Omega}_\beta X(X'X)^{-1},
\end{equation*}
where
\begin{equation*}
\widehat{\Omega}_\beta = \diag\{\hat{e}_t^2 \, g_{t}\},
\end{equation*}
with
\begin{equation*}
g_{t} = \frac{n}{n-p} \left(\frac{1}{F_{\text{Beta}}(w_t; \tilde{a}, \tilde{b})}\right)^{c_1/n^{c_2}}.
\end{equation*}

The constants $c_1$ and $c_2$ are positive tuning parameters that control the
strength of the Beta-based adjustment through the exponent $c_1/n^{c_2}$.
Throughout the paper, they are assumed to be fixed, so that the exponent
decreases monotonically to zero as $n\rightarrow\infty$. Consequently, the
additional Beta-based adjustment gradually vanishes, and the proposed
estimator recovers the asymptotic behavior of the conventional
heteroskedasticity-consistent covariance estimators. The limiting case
$c_1=0$ reduces the proposed adjustment to the HC1 scaling, whereas positive
values of $c_1$ introduce an additional leverage-dependent correction whose
strength increases with $c_1$. Likewise, larger values of $c_2$ accelerate
the decay of the adjustment as the sample size increases. For illustration,
with the recommended values $c_1=7$ and $c_2=0.75$, the exponent equals
approximately $0.74$ for $n=20$, $0.37$ for $n=50$, $0.13$ for $n=200$, and
$0.07$ for $n=500$, providing substantial finite-sample corrections while
ensuring that the additional adjustment becomes asymptotically negligible.

Like tuning parameters commonly used in regularization and smoothing methods, $c_1$ and $c_2$ are not model parameters and no theoretically optimal values are currently available. Instead, they were calibrated empirically through a preliminary Monte Carlo study covering a broad range of sample sizes, heteroskedasticity levels, and leverage configurations. Candidate values were evaluated according to the finite-sample performance of the resulting quasi-$t$ tests, with primary emphasis on empirical size accuracy and confidence interval coverage. The choice $(c_1,c_2)=(7,0.75)$ consistently provided an excellent compromise across the scenarios considered, avoiding both under- and over-correction. Moreover, moderate departures from the recommended values had little impact on the overall inferential performance of the proposed estimator, indicating that the procedure is reasonably insensitive to the precise choice of these constants.

\subsection{Consistency}

The proposed estimator inherits the first-order asymptotic properties of the
classical heteroskedasticity-consistent covariance matrix estimators. Assume the usual fixed-design linear regression framework with fixed $p$, and suppose that
\[
\max_{1 \leq t \leq n} \Vert \bm{x}_t \Vert = o(\sqrt{n}), \,\, \frac{X'X}{n}\rightarrow Q_1,
\,\, \text{ and } \,\,
\frac{X'\Omega X}{n}\rightarrow Q_2,
\]
where $Q_1$ and $Q_2$ are finite positive definite matrices. It follows from the first two assumptions that $\max_{1 \leq t \leq n} h_t \rightarrow 0$.

Because $\Psi_n = (X'X)^{-1}X'\Omega X(X'X)^{-1} = O(n^{-1})$, the exact covariance matrix itself converges to the zero matrix, so convergence of $\widehat\Psi_{\text{HC}_\beta}$ to $\Psi_n$ on this unscaled level would be a degenerate and uninformative statement. The non-degenerate object is the rescaled covariance: writing $A_n = X'X/n \rightarrow Q_1$ and $C_n = X'\Omega X/n \rightarrow Q_2$, we have $n\Psi_n = A_n^{-1} C_n A_n^{-1} \rightarrow \Phi \equiv Q_1^{-1}Q_2Q_1^{-1}$, a finite positive definite matrix. Throughout this subsection, consistency of $\widehat\Psi_{\text{HC}_\beta}$ is understood in this sense, namely
\[
n\widehat\Psi_{\text{HC}_\beta} \xrightarrow{p} \Phi, \qquad \text{equivalently} \qquad n\bigl(\widehat\Psi_{\text{HC}_\beta} - \Psi_n\bigr) \xrightarrow{p} 0.
\]
Establishing this requires identifying the probability limit of the HC0 sandwich meat itself, which is not implied by the two design conditions above. We therefore add the transfer condition
\begin{equation}
\frac{1}{n}X'\widehat\Omega_{\text{HC0}}X \xrightarrow{p} Q_2, \quad \text{where} \quad \widehat\Omega_{\text{HC0}} = \diag\{\hat{e}_t^2\}, \tag{HC0-T}
\end{equation}
imported from the theory of the HC0 estimator rather than implied by the design conditions stated above. Theorem~1(i) in \citet{White_1980} establishes that $\Vert n^{-1}X'\widehat\Omega_{\text{HC0}}X - n^{-1}X'\Omega X\Vert \to 0$ almost surely, so that, combined with the design condition $n^{-1}X'\Omega X \to Q_2$, it delivers (HC0-T) --- indeed in the stronger almost sure mode, which implies the convergence in probability used here. That result, however, is obtained under his Assumptions 1, 2, 3(a) and 4, which require uniformly bounded moments of order slightly greater than one for the squared errors and for the fourth-order products of the regressors; in the fixed-regressor case these requirements amount to uniformly bounded regressors, and they are not implied by $\max_{1 \leq t \leq n} \Vert \bm{x}_t \Vert = o(\sqrt{n})$. We therefore state (HC0-T) as a high-level condition rather than derive it under those stronger moment restrictions: it is the standard hypothesis under which HC0 itself is known to be consistent for $Q_2$, it holds in particular under White's assumptions, and it is this hypothesis that HC$_\beta$ inherits below.

Because the observed values $w_t$ are truncated to the interval $[0.01, 0.99]$ and the regularized shape parameters $\tilde{a}$ and $\tilde{b}$ are bounded below by $\epsilon > 0$ and bounded above by $A_{\max}$ and $B_{\max}$, respectively (with the convention that $\tilde{a} = A_{\max}$ and $\tilde{b} = B_{\max}$ whenever $s_w^2 = 0$), they remain bounded away from zero and infinity uniformly in $n$. Consequently, there exists a constant $\delta \in (0,1)$, independent of $n$, such that
\[
\inf_{1 \le t \le n}
F_{\rm Beta}(w_t;\tilde a,\tilde b)
\ge
\inf_{\substack{w \in [0.01, 0.99] \\ a, b \in [\epsilon, \max(A_{\max}, B_{\max})]}}
F_{\rm Beta}(w; a, b)
\ge\delta > 0.
\]
The infimum on the right-hand side is indeed strictly positive: the map $(w,a,b) \mapsto F_{\rm Beta}(w;a,b)$ is continuous and strictly positive on the compact set $\mathcal{K} = [0.01,0.99] \times [\epsilon, \max(A_{\max}, B_{\max})]^2$, so, by the Weierstrass extreme value theorem, it attains its minimum on $\mathcal{K}$, and one may take $\delta = \min_{\mathcal{K}} F_{\rm Beta}(w;a,b) > 0$. The truncation, the floor $\epsilon$ and the ceilings $A_{\max}$ and $B_{\max}$ place all the arguments in this same deterministic compact set, for every $n$.
Since $0<F_{\text{Beta}}(\cdot)\le1$, it follows that
\begin{equation}\label{E:sup_finite}
\sup_{1 \le t \le n}
\left|
\log
\bigl(
F_{\text{Beta}}(w_t;\tilde a,\tilde b)
\bigr) \right|
\le -\log(\delta)
<\infty.
\end{equation}

Now,
\[
g_t
=
\frac{n}{n-p}
\exp\!\left(
-
\frac{c_1}{n^{c_2}}
\log \!
\left(
F_{\text{Beta}}(w_t;\tilde a,\tilde b)
\right) \right).
\]
Let $z_t = - \frac{c_1}{n^{c_2}} \log \bigl( F_{\text{Beta}}(w_t;\tilde a,\tilde b) \bigr)$.
By \eqref{E:sup_finite}, there exists a constant $M>0$ such that $|z_t| \le M n^{-c_2}$ for all $t$, which implies that $z_t \to 0$ uniformly in $t$ as $n \to \infty$. Moreover, $z_t \ge 0$, since $F_{\text{Beta}}(\cdot) \le 1$ implies $\log \bigl( F_{\text{Beta}}(\cdot) \bigr) \le 0$. Writing $r_n = M n^{-c_2}$, the elementary inequality $\vert e^z - 1 \vert \le z e^z$, valid for $z \ge 0$, then yields, with a constant that is uniform in $t$,
\[
\sup_{1\le t\le n} \left| g_t - \frac{n}{n-p} \right|
= \frac{n}{n-p}\, \sup_{1\le t\le n} \left| e^{z_t} - 1 \right|
\le \frac{n}{n-p}\, r_n e^{r_n}
= O(n^{-c_2}) = o(1).
\]

Since $n/(n-p) - 1 = p/(n-p) = O(n^{-1}) = o(1)$, the triangle inequality yields
\[
\sup_{1\le t\le n}|g_t - 1|
\le
\sup_{1\le t\le n}\left|g_t - \frac{n}{n-p}\right| + \left|\frac{n}{n-p}-1\right|
= O(n^{-c_2}) + O(n^{-1}) = o(1).
\]

Let
\[
\Delta_1 = \frac{1}{n} X' \widehat\Omega_\beta X - \frac1n X'\widehat\Omega_{\text{HC0}} X
\,\, \text{ and } \,\,
\Delta_2 = \widehat\Psi_{\text{HC}_\beta}
-
\widehat\Psi_{\text{HC0}}.
\]
We can express $\Delta_1$ as
\[
\Delta_1 = \frac{1}{n} \sum_{t=1}^n \hat{e}_t^2 (g_t - 1) \bm{x}_t \bm{x}_t'.
\]
Since $\hat{e}_t^2 \ge 0$ and $\bm{x}_t \bm{x}_t'$ is positive semi-definite, each matrix $\hat{e}_t^2 \, \bm{x}_t \bm{x}_t'$ is positive semi-definite. Let $K_n = \sup_{1 \le t \le n} |g_t - 1|$ (which was shown above to satisfy $K_n = o(1)$). Note that, under fixed design, $K_n$ is deterministic, since $g_t$ is a deterministic function of $X$; the randomness in the bound below enters only through $n^{-1}X'\widehat\Omega_{\text{HC0}}X$. Here and in what follows, $\Vert \cdot \Vert$ denotes the spectral norm, so that $\Vert A \Vert = \sup_{\Vert \bm{v} \Vert = 1} \vert \bm{v}'A \bm{v} \vert$ for symmetric $A$; it is this characterization that converts a bound on quadratic forms into a bound on the norm. Bounding the associated quadratic forms, we obtain
\[
\|\Delta_1\| \le K_n \left\| \frac{1}{n} \sum_{t=1}^n \hat{e}_t^2 \bm{x}_t \bm{x}_t' \right\| = \left( \sup_{1 \le t \le n} |g_t - 1| \right) \left\| \frac{1}{n} X' \widehat\Omega_{\text{HC0}} X \right\|.
\]
Condition (HC0-T) implies, in particular, that $n^{-1} X' \widehat\Omega_{\text{HC0}} X = O_p(1)$, and only this weaker stochastic boundedness is needed at the present step: it controls the norm of $\Delta_1$, whereas identifying the limit of $n^{-1}X'\widehat\Omega_{\text{HC0}}X$ as $Q_2$, which is required only in the final step below, uses the full force of (HC0-T). Since $\sup_t |g_t - 1| = o(1)$, it follows that
\[
\|\Delta_1\| = o(1) \times O_p(1) = o_p(1).
\]

Since $(X'X)^{-1} = O(n^{-1})$ and, by $\Delta_1 = o_p(1)$,
\[
X'(\widehat\Omega_\beta-\widehat\Omega_{\text{HC0}}) X = n\Delta_1 = o_p(n).
\]
Writing
$\Delta_2 = (X'X)^{-1} \bigl[ X'(\widehat\Omega_\beta-\widehat\Omega_{\text{HC0}}) X \bigr] (X'X)^{-1}$
and using the submultiplicativity of the spectral norm, together with
$\|(X'X)^{-1}\| = O(n^{-1})$ applied on each side, we obtain
\[
\|\Delta_2\|
\le
\|(X'X)^{-1}\|^2 \, \bigl\| X'(\widehat\Omega_\beta-\widehat\Omega_{\text{HC0}}) X \bigr\|
=
O(n^{-2}) \times o_p(n)
=
o_p(n^{-1}).
\]
Hence, $\text{HC}_\beta$ and HC0 are asymptotically equivalent on the $n^{-1}$ scale: $n\Delta_2 = o_p(1)$. To conclude consistency of $\widehat\Psi_{\text{HC}_\beta}$ itself in the sense defined above, invoke (HC0-T): since $\Delta_1 = o_p(1)$,
\[
\frac{1}{n}X'\widehat\Omega_\beta X = \frac{1}{n}X'\widehat\Omega_{\text{HC0}}X + \Delta_1 \xrightarrow{p} Q_2.
\]
Because $A_n^{-1} \to Q_1^{-1}$, the continuous mapping theorem applied to $(A,B)\mapsto A^{-1}BA^{-1}$ gives
\[
n\widehat\Psi_{\text{HC}_\beta} = A_n^{-1}\left(\frac1n X'\widehat\Omega_\beta X\right)A_n^{-1} \xrightarrow{p} Q_1^{-1}Q_2Q_1^{-1} = \Phi.
\]
Together with $n\Psi_n \to \Phi$, this yields $n\bigl(\widehat\Psi_{\text{HC}_\beta} - \Psi_n\bigr) \xrightarrow{p} 0$, i.e.\ the consistency of $\widehat\Psi_{\text{HC}_\beta}$, transferred conditionally from the consistency of HC0 under (HC0-T).

The consistency argument extends to stochastic regressors under the following explicit conditions. Let $\bm{0}$ denote the $n$-dimensional null vector. Suppose $p$ is fixed, $X$ has full column rank with probability tending to one, $\E(\bm{e}\mid X) = \bm{0}$, $\Var(\bm{e}\mid X) = \Omega_X = \diag\{\sigma_t^2(X)\}$,
\[
\frac{X'X}{n} \xrightarrow{p} Q_1, \,\, \text{ and } \,\,  \frac{X'\Omega_X X}{n} \xrightarrow{p} Q_2,
\]
with $Q_1, Q_2$ positive definite. We further assume a uniform moment condition on the regressors, analogous to Assumption~4 in \citet{White_1980}: there exist finite positive constants $\delta_0$ and $\Delta_0$ such that
\[
\E\bigl(\vert x_{tj}^2 x_{tk} x_{tl}\vert^{1+\delta_0}\bigr) < \Delta_0, \quad j,k,l = 1,\ldots,p,
\]
for all $t$. This moment condition plays, for stochastic regressors, the role that $\max_{1\le t\le n}\Vert \bm{x}_t \Vert = o(\sqrt n)$ plays under fixed design: it rules out regressors with excessively heavy tails, so that a weak law of large numbers applies to the fourth-order sample averages entering $n^{-1}X'\Omega_X X$ and $n^{-1}X'\widehat\Omega_{\text{HC0}}X$. Together with the moment condition, we assume the stochastic-regressor analogue of the transfer condition,
\begin{equation}
\frac{1}{n}X'\widehat\Omega_{\text{HC0}}X \xrightarrow{p} Q_2. \tag{HC0-T$'$}
\end{equation}
Condition (HC0-T$'$) is not made redundant by the moment condition just stated: the matrix that Theorem~1 in \citet{White_1980} compares $n^{-1}X'\widehat\Omega_{\text{HC0}}X$ with is the average of expectations $n^{-1}\sum_{t=1}^n \E\bigl(\sigma_t^2(X)\bm{x}_t\bm{x}_t'\bigr)$, whereas the object appearing in the conditions above is the random matrix $n^{-1}X'\Omega_X X$; bridging the two requires an additional law of large numbers. Stating (HC0-T$'$) directly avoids that detour. Under these conditions, because the transformed leverages are truncated to $[0.01,0.99]$ and the shape parameters are bounded below by $\epsilon>0$ and above by construction, there exists a deterministic constant $M<\infty$ such that
\[
\sup_{1\le t\le n}
\bigl|\log \bigl( F_{\text{Beta}}(w_t;\tilde a,\tilde b) \bigr) \bigr|
\le M.
\]
The truncation, the floor and the ceilings are applied pathwise, so the compact set $\mathcal{K}$, and hence $\delta$ and $M$, depend neither on the sample nor on $n$. The uniform negligibility bound therefore carries over from the fixed-design case without modification and, in fact, in a stronger form: there is a deterministic constant $C<\infty$, depending only on $p$, $c_1$, $c_2$, $\epsilon$, $A_{\max}$ and $B_{\max}$, such that
\[
\sup_{1\le t\le n} |g_t - 1| \le C\, n^{-\min\{c_2,\,1\}}
\]
for every $n \ge 2p$ and every realization of the design, so that the rate is $O\bigl(n^{-\min\{c_2,1\}}\bigr)$ rather than merely $O_p\bigl(n^{-\min\{c_2,1\}}\bigr)$. No moment condition and no condition on the behavior of the leverages is used at this step; this is precisely what the numerical safeguards built into $\text{HC}_\beta$ deliver.

Under (HC0-T$'$), $n^{-1}X'\widehat\Omega_{\text{HC0}}X = O_p(1)$; combined with the deterministic bound on $\sup_{1\le t\le n}|g_t-1|$ just obtained, the same quadratic-form bound used in the fixed-design case gives $\Delta_1 = o_p(1)$. Under the standard condition $n^{-1}X'X \xrightarrow{p} Q_1$, with $Q_1$ positive definite, we also have $(X'X)^{-1} = O_p(n^{-1})$. Combined with $\Delta_1 = o_p(1)$, this yields $\Delta_2 = o_p(n^{-1})$, i.e.\ $n\Delta_2 = o_p(1)$, so that $\text{HC}_\beta$ and HC0 remain asymptotically equivalent on the $n^{-1}$ scale. As in the fixed-design case, (HC0-T$'$) identifies $\text{plim}\, n^{-1}X'\widehat\Omega_{\text{HC0}}X = Q_2$, so that $n^{-1}X'\widehat\Omega_\beta X \xrightarrow{p} Q_2$ and, by the continuous mapping theorem applied to $(A,B)\mapsto A^{-1}BA^{-1}$,
\[
n\widehat\Psi_{\text{HC}_\beta} \xrightarrow{p} Q_1^{-1}Q_2Q_1^{-1},
\]
which is the consistency of $\widehat\Psi_{\text{HC}_\beta}$, transferred conditionally from that of HC0, in the case of stochastic regressors.

\section{Simulation results}\label{S:simulation}

The Monte Carlo simulations were conducted using $10{,}000$ replications and implemented in the statistical computing environment \textsf{R} \citep{R_manual}. The objective is to assess the finite-sample inferential performance of the proposed $\text{HC}_\beta$ estimator in terms of both hypothesis testing and interval estimation. Specifically, we evaluate the finite-sample behavior of quasi-$t$ tests and confidence interval coverage based on the proposed estimator and compare the results with those obtained using the main heteroskedasticity-consistent estimators HC0, HC3, HC4, and HC4m, as well as the classical ordinary least squares estimator. The experiments are conducted under the standard fixed-design linear regression model because the proposed methodology modifies only the leverage-based adjustment applied to the squared residuals. Accordingly, the simulation design focuses on the two components that directly determine the behavior of HC estimators: the degree of heteroskedasticity and the configuration of leverage values. The objective is not to demonstrate universal superiority over existing estimators, but rather to assess whether the proposed adaptive adjustment improves finite-sample inference in settings where leverage and heteroskedasticity jointly affect the performance of conventional HC corrections.

The sample sizes considered are $n=50$, $100$, and $200$. The regressor values generated for $n=50$ are replicated two and four times to construct the samples of sizes $n=100$ and $n=200$, respectively, thereby ensuring that the degree of heteroskedasticity, measured by $\lambda=\max\{ \sigma_t^2 \} /\min \{ \sigma_t^2\}$, remains constant as $n$ increases. Other features that may affect regression inference, such as model misspecification, nonlinear regression functions, response contamination, or outlying observations, constitute distinct statistical problems and are not considered here. Their interaction with the proposed estimator is an interesting topic for future research.

The first simulation scenario considers a linear regression model with two regressors and an intercept, defined as
\begin{equation*}
y_t = \beta_1 + \beta_2 x_{t2} + \beta_3 x_{t3} + e_t, 
\quad t = 1, \ldots, n,
\end{equation*}
where the error terms $e_t$ are independent and normally distributed with mean zero and variance $\sigma_t^2$. The regressor values $x_{t2}$ and $x_{t3}$ are obtained as random draws from the standard normal distribution and the standard lognormal distribution, respectively. To introduce an extreme leverage point, the regressor values of the observation with the highest leverage were multiplied by $3.0$. As previously noted, an observation is commonly regarded as a high-leverage point when its leverage exceeds $3\bar{h} = 3p/n$; in this scenario, $3p/n = 0.18$. The regression coefficients are fixed at $\beta_1 = \beta_2 = \beta_3 = 1$. The variance function is $\sigma_t^2 = \exp(\gamma \, x_{t3})$, with $\gamma \in \{0,\, 0.5349,\, 0.7728\}$, which produces scenarios of homoskedasticity ($\lambda = 1$), moderate heteroskedasticity ($\lambda \approx 15$), and strong heteroskedasticity ($\lambda \approx 50$). We test the null hypothesis $\mathcal{H}_0 \colon \beta_3 = 1$ against the alternative hypothesis $\mathcal{H}_1 \colon \beta_3 \neq 1$.

The second simulation scenario increases the dimensionality of the model to five regressors and an intercept:
\begin{equation*}
y_t = \beta_1 + \beta_2 x_{t2} + \beta_3 x_{t3} + \beta_4 x_{t4} + \beta_5 x_{t5} + \beta_6 x_{t6} + e_t, 
\quad t = 1, \ldots, n.
\end{equation*}
The values of $x_{t2}$, $x_{t3}$, $x_{t4}$, and $x_{t5}$ are generated independently from the standard normal distribution. To produce high-leverage observations, the values of regressor $x_{t6}$ are generated as random draws from a standard lognormal distribution. The regressor values of the observation with the highest leverage are multiplied by $1.3$. The regression coefficients are fixed at $\beta_j = 1$, for $j=1,\ldots,6$. The variance function is $\sigma_t^2 = \exp(\gamma \, x_{t6})$, with $\gamma \in \{0,\, 0.5968,\, 0.9396\}$, leading to homoskedasticity ($\lambda = 1$), moderate heteroskedasticity ($\lambda \approx 12$), and strong heteroskedasticity ($\lambda \approx 50$). The main inferential interest focuses on the coefficient associated with the regressor whose values are drawn from the lognormal distribution. Specifically, we test $\mathcal{H}_0 \colon \beta_6 = 1$ against $\mathcal{H}_1 \colon \beta_6 \neq 1$.

Table~\ref{tab:rej_p3} reports the null rejection rates of the quasi-$t$ tests
for the first scenario. Under homoskedasticity, all tests exhibit rejection
rates close to the nominal level. Under heteroskedasticity, the test based on
the OLS estimator becomes severely liberal, incorrectly rejecting
$\mathcal{H}_0$ in more than $30\%$ of the cases when $\lambda \approx 50$.
The tests based on the HC3, HC4, and HC4m estimators display more controlled
size distortions, which decrease slowly as the sample size increases.

Overall, the test based on $\text{HC}_\beta$ delivers the smallest size distortions, although it is slightly conservative in some cases. This result indicates that the proposed adaptive correction provides a favorable balance between controlling size distortions and avoiding excessive rejection rates in finite samples. Under strong heteroskedasticity ($\lambda \approx 50$) with $n=50$, the size of the $\text{HC}_\beta$ test is $6.3\%$, compared with $7.5\%$ for the HC3 test and $7.3\%$ and $6.9\%$ for the HC4 and HC4m tests. Under moderate heteroskedasticity ($\lambda \approx 15$), the test based on the estimator proposed in this paper yields a rejection rate of $6.0\%$, whereas the HC3, HC4, and HC4m tests produce rejection rates of $7.1\%$, $6.9\%$, and $6.5\%$, respectively.

A notable feature of the results is the relatively rapid approach of the rejection rate of the $\text{HC}_\beta$ test to the nominal level as the sample size increases. When the sample size increases to $n = 100$ under $\lambda \approx 50$, the rejection rate of the $\text{HC}_\beta$ test already reaches $4.8\%$, whereas the HC3 and HC4 tests still exhibit rejection rates of $6.8\%$ and $6.8\%$, respectively. For $n = 200$, the $\text{HC}_\beta$ rejection rates lie between $4.0\%$ and $4.5\%$, indicating a mildly conservative behavior that helps prevent excessive rejection in finite samples while remaining close to the nominal significance level.

\begin{table}[htb]
\centering
\begin{threeparttable} 
\caption{Null rejection rates (\%) of quasi-$t$ tests at the 5\% significance level, first simulation scenario.}
\label{tab:rej_p3}
\begin{tabular}{@{}c c *{6}{r}@{}}
\toprule
$\lambda$ & $n$ & OLS & HC0 & HC3 & HC4 & HC4m & $\text{HC}_\beta$ \\
\midrule
\multirow{3}{*}{$1.00$}  & 50  & 5.4 & 7.5 & 5.1 & 4.7 & 4.8 & 4.2 \\
                         & 100 & 5.3 & 6.4 & 5.3 & 5.3 & 5.2 & 3.9 \\
                         & 200 & 5.2 & 5.7 & 5.1 & 5.0 & 5.0 & 4.0 \\
\midrule
\multirow{3}{*}{$\approx 15.00$} & 50  & 23.8 & 10.4 & 7.1 & 6.9 & 6.5 & 6.0 \\
                         & 100 & 22.8 & 7.9 & 6.5 & 6.4 & 6.0 & 4.7 \\
                         & 200 & 22.3 & 6.6 & 5.9 & 5.9 & 5.7 & 4.4 \\
\midrule
\multirow{3}{*}{$\approx 50.00$} & 50  & 31.3 & 11.3 & 7.5 & 7.3 & 6.9 & 6.3 \\
                         & 100 & 29.7 & 8.3 & 6.8 & 6.8 & 6.5 & 4.8 \\
                         & 200 & 29.0 & 6.8 & 6.0 & 6.0 & 5.8 & 4.5 \\
\bottomrule
\end{tabular}
\end{threeparttable} 
\end{table}

Table~\ref{tab:rej_beta6_indep} presents the simulation results for the second scenario. The higher dimensionality of the model makes this setting more challenging, as the larger number of regressors is combined with heterogeneous leverage patterns and increasing levels of heteroskedasticity. Under strong heteroskedasticity ($\lambda \approx 50$) and when $n$ is not large, the tests based on the traditional HC estimators display liberal behavior. With $n=50$, the null rejection rates of the HC0, HC3, HC4, and HC4m tests range from $6.9\%$ (HC4m) to $14.5\%$ (HC0), whereas the $\text{HC}_\beta$ test yields a rejection rate of $6.1\%$, which is the closest to the nominal level among the considered HC-based procedures. As expected, the OLS-based test is severely liberal, rejecting the null hypothesis in $30.3\%$ of the replications.

As the sample size increases, the rejection frequency of $\mathcal{H}_0$ for the $\text{HC}_\beta$ test converges rapidly toward the nominal significance level. For $n = 100$, the null rejection rate of the $\text{HC}_\beta$ test already reaches $4.6\%$, whereas the corresponding rates for HC0 and HC3/HC4 remain noticeably higher at $10.1\%$ and $7.0\%$--$7.1\%$, respectively. In larger samples ($n = 200$), the rejection rates of the test based on the estimator proposed in this paper lie slightly below the nominal level ($4.1\%$--$4.2\%$), showing the mild conservative behavior expected from a procedure that avoids excessive rejection in finite samples. Importantly, this behavior is accompanied by convergence toward the nominal level as the sample size grows.

\begin{table}[htb]
\centering
\begin{threeparttable} 
\caption{Null rejection rates (\%) of quasi-$t$ tests at the 5\% significance level, second simulation scenario.}
\label{tab:rej_beta6_indep}
\begin{tabular}{@{}c c *{6}{r}@{}}
\toprule
$\lambda$ & $n$ & OLS & HC0 & HC3 & HC4 & HC4m & $\text{HC}_\beta$ \\
\midrule
\multirow{3}{*}{$1.00$}  & 50  & 5.7 & 9.1 & 4.9 & 5.6 & 4.6 & 3.9 \\
                         & 100 & 5.5 & 7.2 & 5.2 & 5.6 & 4.9 & 3.6 \\
                         & 200 & 5.4 & 6.2 & 5.3 & 5.4 & 5.1 & 4.1 \\
\midrule
\multirow{3}{*}{$\approx 12.00$} & 50  & 20.3 & 12.9 & 7.2 & 7.6 & 6.3 & 5.5 \\
                         & 100 & 20.4 & 9.3 & 6.5 & 6.8 & 6.0 & 4.4 \\
                         & 200 & 19.5 & 7.2 & 5.8 & 6.0 & 5.6 & 4.2 \\
\midrule
\multirow{3}{*}{$\approx 50.00$} & 50  & 30.3 & 14.5 & 8.0 & 8.2 & 6.9 & 6.1 \\
                         & 100 & 29.7 & 10.1 & 7.0 & 7.1 & 6.4 & 4.6 \\
                         & 200 & 29.0 & 7.5 & 6.1 & 6.2 & 5.9 & 4.1 \\
\bottomrule
\end{tabular}
\end{threeparttable} 
\end{table}

Table~\ref{tab:power_beta6_indep} reports the non-null rejection rates (powers) of the tests based on heteroskedasticity-consistent estimators. For brevity, we focus only on the second simulation scenario. The data are generated with $\beta_6 = 1.8$. Since the tests tend to exhibit liberal behavior, these rejection rates were computed using exact critical values obtained from the size simulations.

Overall, the results show the expected patterns: power increases with the sample size, decreases as the degree of heteroskedasticity intensifies, and remains broadly similar across the competing tests. Under strong heteroskedasticity ($\lambda \approx 50$) and $n=100$, the powers of the tests are around $34\%$, exceeding $60\%$ when the sample size increases to $n=200$.

\begin{table}[htb]
\centering
\begin{threeparttable} 
\caption{Non-null rejection rates (\%) of quasi-$t$ tests at the 5\% significance level, second simulation scenario.}
\label{tab:power_beta6_indep}
\begin{tabular}{@{}c c *{5}{r}@{}}
\toprule
$\lambda$ & $n$ & HC0 & HC3 & HC4 & HC4m & $\text{HC}_\beta$ \\
\midrule
\multirow{3}{*}{$1.00$}  & 50  & 100.0 & 100.0 & 100.0 & 100.0 & 100.0 \\
                         & 100 & 100.0 & 100.0 & 100.0 & 100.0 & 100.0 \\
                         & 200 & 100.0 & 100.0 & 100.0 & 100.0 & 100.0 \\
\midrule
\multirow{3}{*}{$\approx 12.00$} & 50  & 54.4 & 53.2 & 53.6 & 53.2 & 53.4 \\
                         & 100 & 85.1 & 84.7 & 84.4 & 84.4 & 83.9 \\
                         & 200 & 99.1 & 99.1 & 99.1 & 99.1 & 99.1 \\
\midrule
\multirow{3}{*}{$\approx 50.00$} & 50  & 20.0 & 20.3 & 20.2 & 20.3 & 20.4 \\
                         & 100 & 34.4 & 34.2 & 34.2 & 34.1 & 34.0 \\
                         & 200 & 62.0 & 61.9 & 62.0 & 62.0 & 61.8 \\
\bottomrule
\end{tabular}
\end{threeparttable} 
\end{table}

Next, we shift the focus from hypothesis testing to interval estimation. For
brevity, we again consider only the second simulation scenario, a nominal
coverage of $95\%$, and interval estimation for $\beta_6$. The empirical
coverage rates of the different confidence intervals are reported in
Table~\ref{tab:cover_beta6_indep}.

Under strong heteroskedasticity ($\lambda \approx 50$) and a small sample size ($n = 50$), the confidence intervals based on traditional estimators exhibit clear undercoverage. The coverage rates of the confidence intervals based on heteroskedasticity-consistent estimators of $\Var(\hat{\beta}_6)$ range from $85.5\%$ (HC0) to $93.1\%$ (HC4m). In contrast, the $\text{HC}_\beta$ interval attains a coverage rate of $93.9\%$, the closest to the nominal $95\%$ level among the HC estimators considered. The OLS interval shows severe undercoverage ($69.7\%$).

The same pattern persists as the sample size increases. For $n = 100$ under strong heteroskedasticity, the empirical coverage rate of the $\text{HC}_\beta$ confidence interval ($95.4\%$) essentially matches the nominal level, whereas the corresponding coverage rates for HC0, HC3, HC4, and HC4m are $89.9\%$, $93.0\%$, $92.9\%$, and $93.6\%$, respectively. For $n = 200$, all heteroskedasticity-consistent intervals improve, reflecting their common asymptotic consistency. Even so, the proposed estimator continues to produce coverage closest to the nominal level, indicating that the Beta-based adjustment remains beneficial beyond very small samples.

\begin{table}[htb]
\centering
\begin{threeparttable} 
\caption{Empirical coverage rates (\%) of $95\%$ confidence intervals for $\beta_6$, second simulation scenario.}
\label{tab:cover_beta6_indep}
\begin{tabular}{@{}c c *{6}{r}@{}}
\toprule
$\lambda$ & $n$ & OLS & HC0 & HC3 & HC4 & HC4m & $\text{HC}_\beta$ \\
\midrule
\multirow{3}{*}{$1.00$}  & 50  & 94.3 & 91.0 & 95.1 & 94.4 & 95.4 & 96.2 \\
                         & 100 & 94.6 & 92.8 & 94.8 & 94.4 & 95.1 & 96.4 \\
                         & 200 & 94.6 & 93.8 & 94.8 & 94.7 & 94.9 & 96.0 \\
\midrule
\multirow{3}{*}{$\approx 12.00$} & 50  & 79.7 & 87.1 & 92.9 & 92.4 & 93.7 & 94.5 \\
                         & 100 & 79.6 & 90.7 & 93.5 & 93.2 & 94.0 & 95.6 \\
                         & 200 & 80.5 & 92.8 & 94.2 & 94.0 & 94.5 & 95.9 \\
\midrule
\multirow{3}{*}{$\approx 50.00$} & 50  & 69.7 & 85.5 & 92.0 & 91.8 & 93.1 & 93.9 \\
                         & 100 & 70.4 & 89.9 & 93.0 & 92.9 & 93.6 & 95.4 \\
                         & 200 & 71.1 & 92.5 & 93.9 & 93.8 & 94.1 & 95.9 \\
\bottomrule
\end{tabular}
\end{threeparttable} 
\end{table}

\section{Empirical applications}\label{S:applications}

We now turn to empirical applications that illustrate the estimation of the covariance matrix of $\bm{\hat{\beta}}$ in HC-form. We consider the estimator proposed in this paper ($\text{HC}_\beta$) and, for reference, the alternative estimators HC0, HC3, HC4, and HC4m. These applications aim to assess whether the $\text{HC}_\beta$ estimator yields covariance estimates that are comparable to, or potentially more accurate than, those obtained from conventional heteroskedasticity-consistent methods.

In addition to these estimators, we also computed HC5 \citep{Cribari+Souza+Vasconcellos_2007} and HC5m \citep{Li+et-al_2016} estimates for all empirical applications; the corresponding results are omitted for brevity. In settings characterized by moderate leverage, HC5 and HC5m generally yield results similar to those obtained with HC4. Under extreme leverage, however, both estimators exhibit an even more pronounced overshooting behavior than HC4, generating substantially larger adjustment factors $g_t$ for highly leveraged observations. This pattern further underscores the practical relevance of the proposed $\text{HC}_\beta$ estimator, whose adaptive construction was specifically designed to prevent excessive growth of the adjustment factors while remaining responsive to the leverage structure of the data.

For comparison, we also report results based on bootstrap methods. Specifically, we consider the pairs bootstrap, denoted by $\text{boot}_p$. For details on bootstrap resampling under heteroskedasticity, see \citet{Flachaire_2005}. All reported bootstrap results are based on 5{,}000 bootstrap replications.

The datasets used in the first three empirical applications are distributed with the \textsl{hcinfer} package for the \textsf{R} statistical computing environment, allowing the corresponding analyses to be reproduced directly from the package.

\subsection{Per capita spending on public schools}

As a starting point, we revisit the dataset modeled by \citet{Cribari+Ferrari+Cordeiro_2000}, \citet{Cribari_2004}, \citet{Cribari+Pereira_2019}, and \citet{Marinho+Cribari+Tomazella_2025}. The dependent variable ($y$) is per capita spending on public schools, while the explanatory variables, $x$ and $x^2$, correspond to per capita income by state in 1979 in the United States and its square. Income is scaled by $10^{-4}$ for numerical convenience. Wisconsin was excluded due to missing data, and the District of Columbia was included, yielding a total of $n = 50$ observations. The regression model considered is
\begin{equation*}
y_t = \beta_1 + \beta_2 x_t + \beta_3 x_t^2 + e_t, \quad t = 1, \ldots, 50.
\end{equation*}
The OLS estimates of the regression coefficients are $\hat{\beta}_1 = 832.9144$, $\hat{\beta}_2 = -1834.2029$, and $\hat{\beta}_3 = 1587.0423$.

The interest lies in testing $\mathcal{H}_0\colon \beta_3 = 0$ versus $\mathcal{H}_1\colon \beta_3 \neq 0$. Under the null hypothesis, the relationship between per capita spending on public schools and per capita income is linear, whereas under the alternative it is quadratic. The standard $z$ test (based on the OLS standard error) yields a test statistic of 3.0574 ($p\text{-value} = 0.0022$), leading to the rejection of the null hypothesis at the 1\% significance level.

There are three observations with leverage values exceeding $3p/n = 0.1800$, namely Alaska ($h_2 = 0.6508$), Mississippi ($h_{24} = 0.2000$), and the District of Columbia ($h_{48} = 0.2079$). In particular, Alaska stands out as a highly influential leverage point. Indeed, the scatterplot displays a reasonably linear pattern except for Alaska.

The HC0, HC3, HC4, HC4m, and bootstrap standard errors of $\hat{\beta}_3$ are, respectively, 829.9927, 1995.2420, 5488.9292, 2553.3270, and 1127.4865. Using these standard errors to compute quasi-$t$ statistics yields the following $p$-values: 0.0558, 0.4264, 0.7724, 0.5342, and 0.1593. The standard error obtained from the proposed $\text{HC}_\beta$ estimator is 1547.4583, with an associated $p$-value of 0.3051. Hence, the quasi-$t$ test based on HC0 rejects the null hypothesis at the 10\% significance level and is marginally close to rejection at the 5\% level ($p\text{-value} = 0.0558$), whereas tests based on the other estimators fail to reject it even at the 10\% level. This suggests that HC0 underestimates the variance of $\hat{\beta}_3$, likely due to the strong leverage exerted by Alaska in the data. The influence of this single observation also explains the result obtained with the pairs bootstrap. Because the pairs bootstrap resamples observations with replacement, Alaska is included at least once in a bootstrap sample with probability $1-(1-1/50)^{50}=0.6358$. Indeed, among the $5{,}000$ bootstrap samples, the $63.92\%$ that include Alaska yield OLS estimates of $\beta_3$ with mean $1867.0325$ and standard deviation $470.0978$, whereas the remaining samples yield mean $-5.0160$ and standard deviation $944.2633$. Consequently, the bootstrap standard error of $1127.4865$ largely reflects the separation between the coefficient estimates obtained from bootstrap samples with and without Alaska, rather than the sampling variability of the estimator under a fixed-design framework. This example illustrates why agreement with the pairs bootstrap should not be interpreted as evidence that a particular HC estimator is necessarily providing the most appropriate standard error.

Since Alaska is a pronounced leverage point, it is instructive to examine its impact on model inference. We removed Alaska (observation 2) from the dataset, re-estimated the parameters, and recomputed the standard errors and quasi-$t$ test $p$-values. The parameter estimates of $\beta_1$, $\beta_2$, and $\beta_3$ became $-209.0336$, $1000.5344$, and $-314.1385$, respectively. This represents a substantial change, especially in $\hat{\beta}_3$, which shifted from $1587.0423$ to $-314.1385$. Hence, the apparent quadratic pattern in the relationship between the response and the regressor is driven almost entirely by a single observation. The HC0, HC3, HC4, HC4m, bootstrap, and $\text{HC}_\beta$ standard errors for $\hat{\beta}_3$ are now $626.6843$, $1103.0287$, $2320.8289$, $1312.0760$, $942.3695$, and $1066.7167$, respectively. All corresponding $p$-values exceed 0.6, so the null hypothesis is not rejected.

\begin{figure}[htb]
    \centering
    \begin{tabular}{cc}

        \subcaptionbox{\centering}{\includegraphics[width=0.45\textwidth]{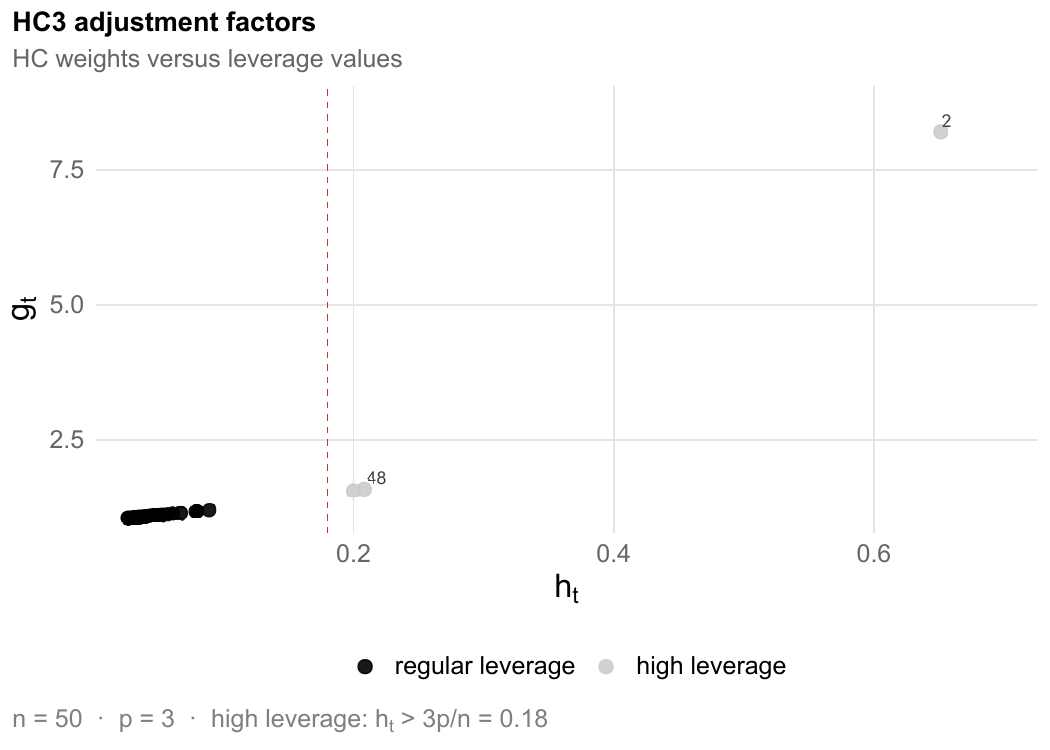}\label{fig:sub_a_01}} &
        \subcaptionbox{\centering}{\includegraphics[width=0.45\textwidth]{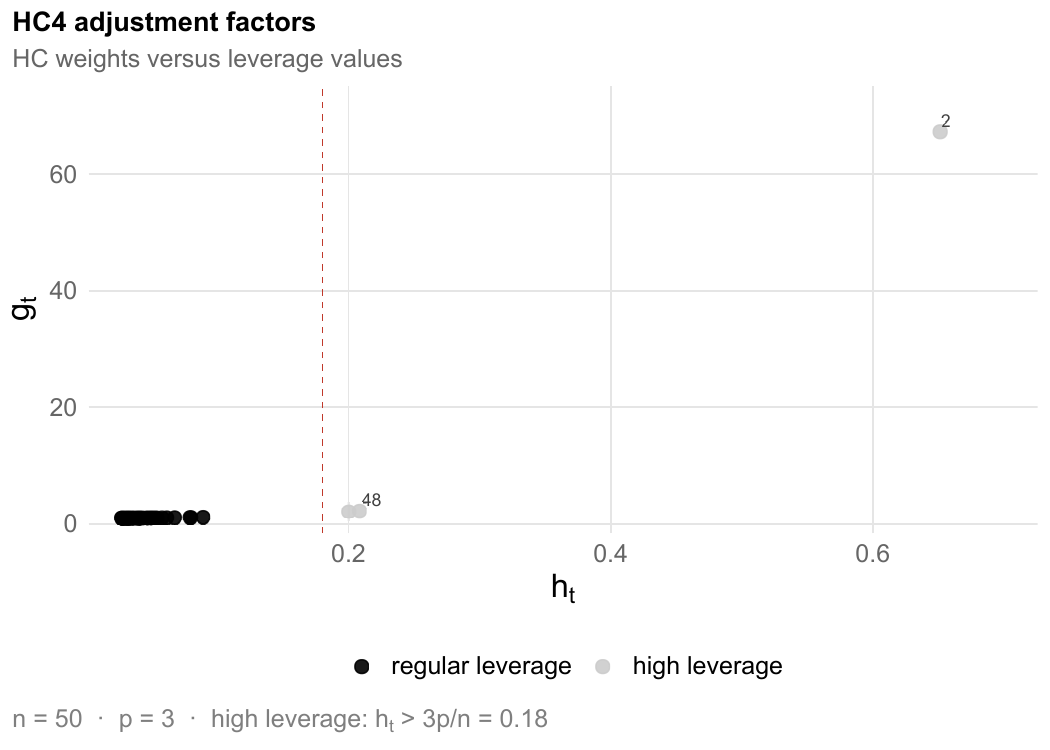}\label{fig:sub_b_01}} \\[1em]

        \subcaptionbox{\centering}{\includegraphics[width=0.45\textwidth]{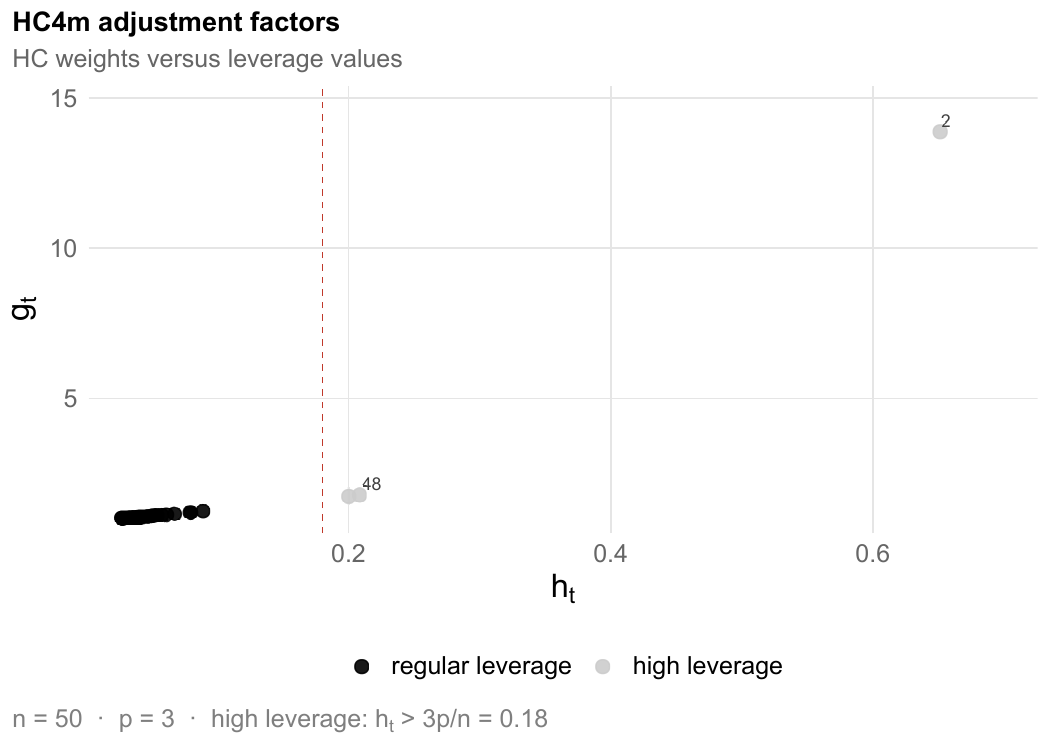}\label{fig:sub_c_01}} &
        \subcaptionbox{\centering}{\includegraphics[width=0.45\textwidth]{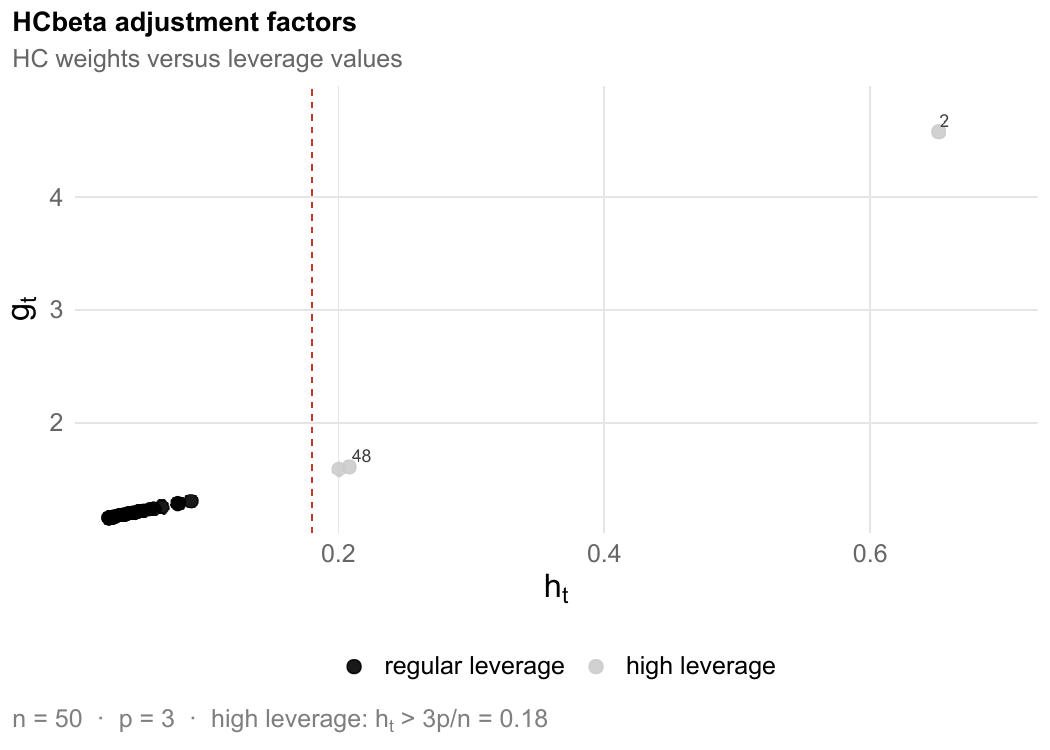}\label{fig:sub_d_01}}

    \end{tabular}

    \caption{Adjustment factors ($g_t$) versus leverages ($h_t$) for HC3 (top-left panel, A), HC4 (top-right panel, B), HC4m (bottom-left panel, C), and $\text{HC}_\beta$ (bottom-right panel, D), complete data; public schools dataset.} 
    \label{F:leverages_01}
\end{figure}

Particularly noteworthy is the magnitude of the HC4 standard error relative to HC3 and, especially, to $\text{HC}_\beta$, when Alaska is included in the sample. Even HC3 and HC4m exceed $\text{HC}_\beta$ considerably. The estimator proposed in this paper produces a noticeably larger standard error than HC0, yet in a controlled manner, without overshooting. To explore this point, Figure~\ref{F:leverages_01} displays the adjustment factors ($g_t$ versus $h_t$) used in the HC3 (top-left panel), HC4 (top-right panel), HC4m (bottom-left panel), and $\text{HC}_\beta$ (bottom-right panel) estimators (complete data). The dashed vertical lines indicate the conventional threshold $3p/n$ used for the identification of high-leverage observations. These plots clearly reveal the overshooting behavior of the HC3, HC4, and HC4m correction factors under strong leverage (i.e.\ for Alaska). In contrast, the growth of $g_t$ with respect to $h_t$ in $\text{HC}_\beta$ is more stable and gradual, avoiding explosive behavior even when $h_t$ is very large. This stability is reflected in the maximal $g_t$ values: 8.2009 (HC3), 67.255 (HC4), 13.878 (HC4m), and 4.5807 ($\text{HC}_\beta$).

\begin{figure}[htb]
\includegraphics[width=0.6\linewidth]{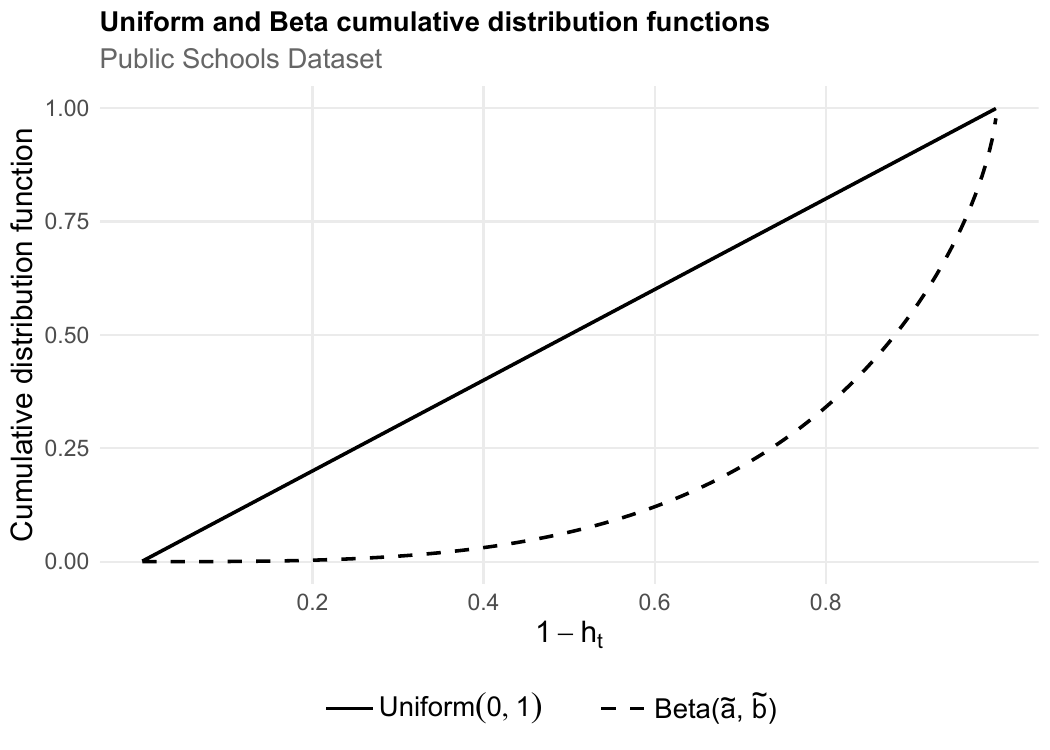}
\caption{Uniform(0,1) and Beta($\tilde{a}, \tilde{b}$) cumulative distribution functions, complete data; public schools dataset.}
\label{F:cumulative_01}
\end{figure}

The estimated Beta parameters (complete data) are $\tilde{a} = 3.1472$ and $\tilde{b} = 0.6690$, which deviate substantially from the uniform case ($a = 1$, $b = 1$). Figure~\ref{F:cumulative_01} displays the Uniform$(0,1)$ --- that is, Beta$(1,1)$ --- and Beta$(\tilde{a}, \tilde{b})$ cumulative distribution functions. Recall that low values of $1 - h_t$ correspond to strong leverage. Consider, for example, $1 - h_t = 0.35$, which is approximately the value obtained for Alaska. Under HC3, we have $g_t = 1 / 0.35^2 = 8.1633$, whereas under $\text{HC}_\beta$, $g_t = (50 / 47) \times 1 / [F(0.35;3.1472, 0.6690)^{0.3723}] = 4.5679$.

Importantly, the fitted Beta density $f(1 - h_t; \tilde{a}, \tilde{b})$ takes values that are markedly smaller than $1$ (the density of the standard uniform distribution) under strong leverage (large $h_t$, small $1 - h_t$) and markedly larger under weak leverage (small $h_t$, large $1 - h_t$). This behavior induces a smoother and more gradual adjustment of the squared residuals as the degree of leverage varies. In other words, the $\text{HC}_\beta$ estimator responds adaptively to the empirical leverage distribution, tempering the excessive inflation observed in traditional correction schemes while still providing adequate protection against downward bias. This adaptive, distribution-driven mechanism explains the enhanced stability of $\text{HC}_\beta$ in the presence of highly influential observations.

\subsection{Boston house prices}

We use data collected from the real estate section of the Boston Globe in 1990, which record the sale prices of homes in the Boston, MA area. The response variable ($y$) is the house price (in thousands of U.S.\ dollars), and the explanatory variables are as follows: $x_2$ represents the lot size (in square feet), $x_3$ the number of bedrooms, and $x_4$ the house size (in square feet). The sample size is $n = 88$. The regression model is
\begin{equation*}
y_t = \beta_1 + \beta_2 x_{t2} + \beta_3 x_{t3} + \beta_4 (x_{t3} \!\times\! x_{t4}) + e_t, \quad t = 1, \ldots, 88,
\end{equation*}
where the interaction term allows the effect of the number of bedrooms to vary with the size of the house. The ordinary least squares estimates of the regression coefficients are $\hat{\beta}_1 = 231.8807$, $\hat{\beta}_2 = 0.0020$, $\hat{\beta}_3 = -48.8613$, and $\hat{\beta}_4 = 0.0293$.

There are five observations with leverage values exceeding $3p/n = 0.1364$: cases $29$ ($h_{29} = 0.2696$), $38$ ($h_{38} = 0.1617$), $63$ ($h_{63} = 0.2768$), $73$ ($h_{73} = 0.1555$), and $77$ ($h_{77} = 0.8517$). In particular, observation $77$ exhibits extremely high leverage. This corresponds to a house with a lot size of $92{,}681$ square feet --- more than ten times larger than the sample average of $9{,}019.86$ square feet.

The OLS, HC0, HC3, HC4, HC4m, bootstrap, and $\text{HC}_\beta$ standard errors of $\hat{\beta}_1, \ldots, \hat{\beta}_4$ are reported in Table~\ref{T:hp_se_01}. For the $\text{HC}_\beta$ estimator, the fitted Beta parameters are $\tilde{a} = 2.5048$ and $\tilde{b} = 0.4643$. Inspection of Table~\ref{T:hp_se_01} reveals that the $\text{HC}_\beta$ standard errors are, as expected given the presence of strong leverage, moderately inflated relative to those from HC0. However, this inflation is considerably smoother and less aggressive than that observed for HC3, HC4m, and especially HC4. For instance, the standard error of $\hat{\beta}_3$ increases from $17.0639$ (HC0) to $21.4135$ ($\text{HC}_\beta$), a rise of just over 25\%. By contrast, the corresponding values for HC3, HC4, and HC4m are $33.1596$ (a $94\%$ increase), $172.4775$ (a $910\%$ increase), and $46.7080$ (a $174\%$ increase), respectively. These exaggerated adjustments illustrate the overshooting problem typical of traditional high-order HC estimators under strong leverage. In contrast, the $\text{HC}_\beta$ estimator inflates squared residuals in a more controlled and data-driven manner, delivering enhanced stability without excessive correction. It is also noteworthy that the $\text{HC}_\beta$ standard error ($21.4135$) is very close to its pairs bootstrap counterpart ($20.7492$), without exhibiting the substantial discrepancies observed relative to HC3, HC4, and HC4m. The same qualitative pattern holds for the remaining regression coefficients reported in the table, with the pairs bootstrap estimates consistently much closer to those obtained from $\text{HC}_\beta$ than to the corresponding HC3, HC4, and HC4m estimates.

\begin{table}[htb]
    \centering
    \begin{threeparttable}
    \caption{Standard errors with complete data; house prices dataset.}
    \label{T:hp_se_01}
    \begin{tabular}{@{}l *{7}{r}@{}}
        \toprule
        & \multicolumn{1}{c}{OLS} & \multicolumn{1}{c}{HC0} & \multicolumn{1}{c}{HC3} & \multicolumn{1}{c}{HC4} & \multicolumn{1}{c}{HC4m} & \multicolumn{1}{c}{$\text{HC}_\beta$} & \multicolumn{1}{c}{$\text{boot}_p$} \\
        \midrule
        $\text{se}(\hat{\beta}_1)$ & 32.3163 & 40.3954 & 93.5218 & 535.5203 & 137.9477 & 52.5037 & 54.2481 \\
        $\text{se}(\hat{\beta}_2)$ & 0.0006 & 0.0011 & 0.0067 & 0.0451 & 0.0108 & 0.0021 & 0.0035 \\
        $\text{se}(\hat{\beta}_3)$ & 13.2230 & 17.0639 & 33.1596 & 172.4775 & 46.7080 & 21.4135 & 20.7492 \\
        $\text{se}(\hat{\beta}_4)$ & 0.0030 & 0.0046 & 0.0102 & 0.0582 & 0.0150 & 0.0059 & 0.0062 \\
        \bottomrule
    \end{tabular}
    \end{threeparttable}
\end{table}

Consider the test of $\mathcal{H}_0\colon \beta_3 = 0$ against $\mathcal{H}_1\colon \beta_3 \neq 0$. Table~\ref{tab:tests_pvalues_x3_02} reports the $p$-values of the quasi-$t$ tests obtained using the full dataset and after removing observations $63$ and $77$, which exhibit the two largest leverage values. When all observations are included, only the tests based on the HC0, bootstrap, and $\text{HC}_\beta$ standard errors lead to rejection of the null hypothesis at conventional significance levels: HC0 rejects at the 1\% level, while pairs bootstrap and $\text{HC}_\beta$ reject at the 5\% level. The standard errors from HC3, HC4, and HC4m appear excessively inflated due to strong leverage, which in turn causes the corresponding quasi-$t$ statistics to be too small in absolute value. Interestingly, when the two most influential points (observations $63$ and $77$) are removed, all tests yield $p$-values below 0.01, indicating strong evidence against the null hypothesis. The $\text{HC}_\beta$ test therefore provides coherent inference in both scenarios, unlike HC3, HC4, and HC4m, which exhibit instability under high leverage.

A similar pattern emerges when testing $\mathcal{H}_0\colon \beta_2 = 0$ against $\mathcal{H}_1\colon \beta_2 \neq 0$. The corresponding $p$-values are presented in Table~\ref{tab:tests_pvalues_x2_02}. When all observations are used, none of the tests reject $\mathcal{H}_0$ at the 5\% significance level, although HC0 rejects at 10\%. Notably, the $p$-value from the $\text{HC}_\beta$ test is considerably smaller than those obtained from HC3, HC4, HC4m, and pairs bootstrap. When the two high-leverage observations are excluded, all tests reject $\mathcal{H}_0$ at the 1\% level, revealing that these leverage points obscure the true significance of regressor $x_2$. Once again, we observe the typical overshooting behavior of HC3, HC4, and HC4m under strong leverage, whereas the $\text{HC}_\beta$ estimator delivers a more stable standard error --- correcting HC0’s downward bias without resorting to excessive inflation.

\begin{table}[htb]
\centering
\begin{threeparttable}
\caption{Tests' $p$-values, $\mathcal{H}_0: \beta_3 = 0$; house prices dataset.}
\label{tab:tests_pvalues_x3_02}
\begin{tabular}{@{}l r l r@{}}
\toprule
\multicolumn{2}{c}{complete data, $n = 88$} & \multicolumn{2}{c}{incomplete data, $n = 86$} \\
\cmidrule(lr){1-2} \cmidrule(lr){3-4}
Test & $p$-value & Test & $p$-value \\
\midrule
HC0           & 0.0042 & HC0           & 0.0023 \\
HC3           & 0.1406 & HC3           & 0.0058 \\
HC4           & 0.7770 & HC4           & 0.0093 \\
HC4m          & 0.2955 & HC4m          & 0.0068 \\
$\text{HC}_\beta$  & 0.0225 & $\text{HC}_\beta$  & 0.0093 \\
$\text{boot}_p$ & 0.0185 & $\text{boot}_p$ & 0.0092 \\
\bottomrule
\end{tabular}
\end{threeparttable}
\end{table}

\begin{table}[htb]
\centering
\begin{threeparttable}
\caption{Tests' $p$-values, $\mathcal{H}_0: \beta_2 = 0$; house prices dataset.}
\label{tab:tests_pvalues_x2_02}
\begin{tabular}{@{}l r l r@{}}
\toprule
\multicolumn{2}{c}{complete data, $n = 88$} & \multicolumn{2}{c}{incomplete data, $n = 86$} \\
\cmidrule(lr){1-2} \cmidrule(lr){3-4}
Test & $p$-value & Test & $p$-value \\
\midrule
HC0           & 0.0673 & HC0           & $<$0.0001 \\
HC3           & 0.7667 & HC3           & 0.0004 \\
HC4           & 0.9648 & HC4           & 0.0046 \\
HC4m          & 0.8537 & HC4m          & 0.0008 \\
$\text{HC}_\beta$  & 0.3540 & $\text{HC}_\beta$  & 0.0009 \\
$\text{boot}_p$ & 0.5700 & $\text{boot}_p$ & $<$0.0001 \\
\bottomrule
\end{tabular}
\end{threeparttable}
\end{table}

\begin{figure}[htb]
    \centering
    \begin{tabular}{cc}
        \subcaptionbox{\centering}{\includegraphics[width=0.45\textwidth]{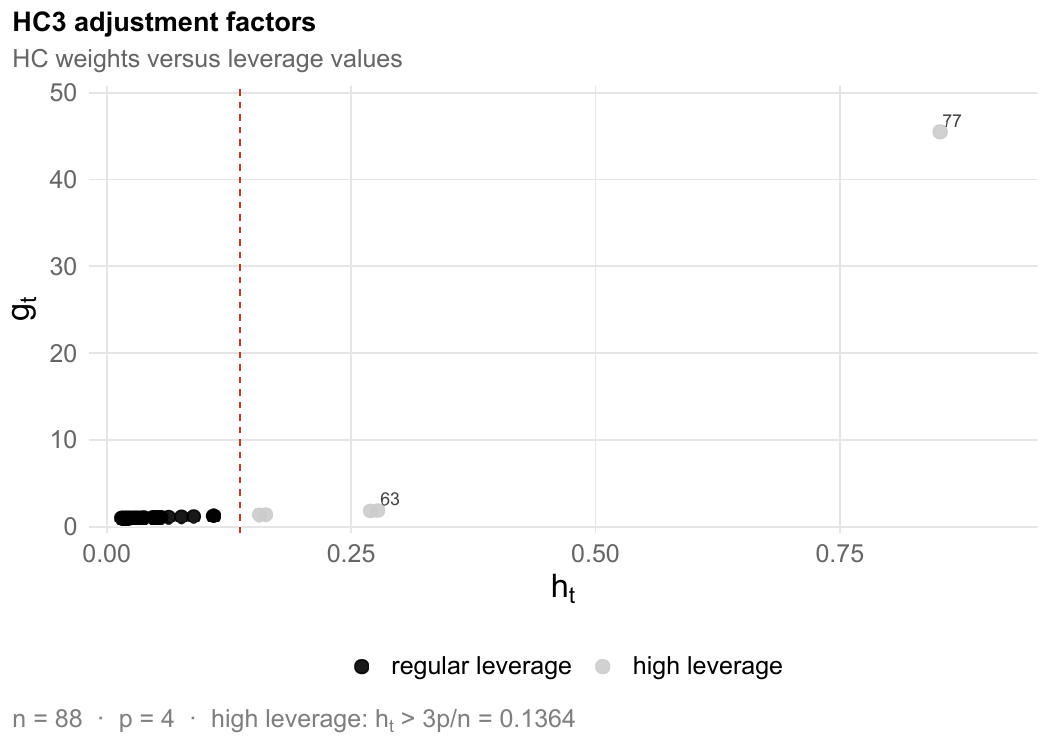}\label{fig:sub_a_02}} &
        \subcaptionbox{\centering}{\includegraphics[width=0.45\textwidth]{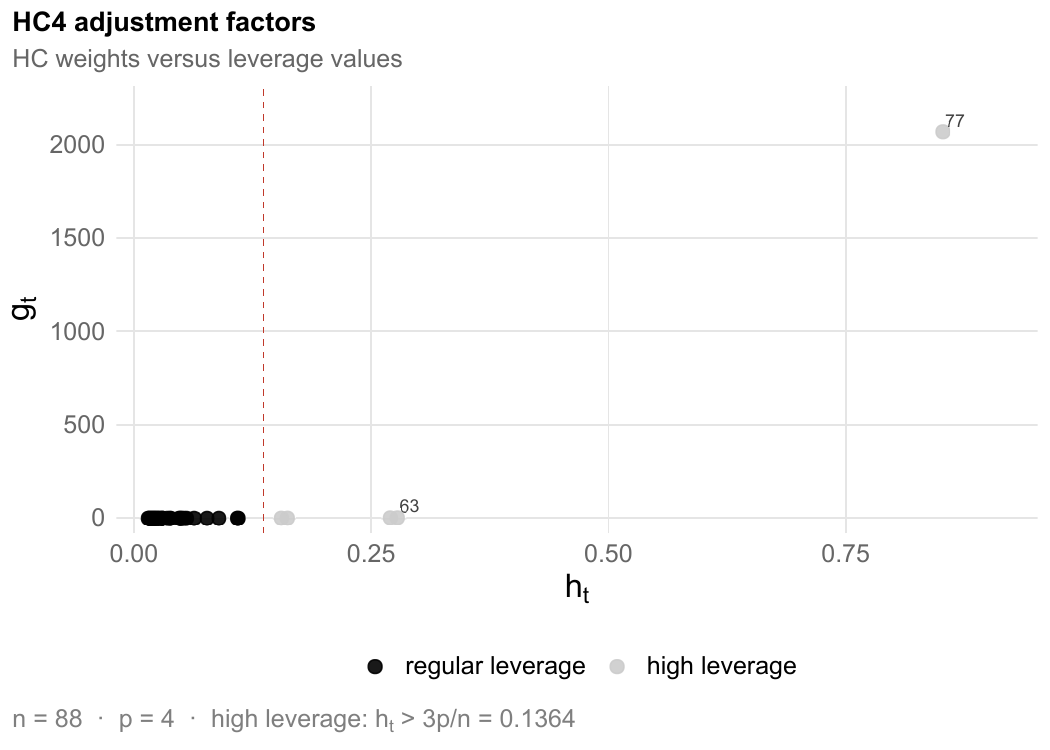}\label{fig:sub_b_02}} \\[1em]

        \subcaptionbox{\centering}{\includegraphics[width=0.45\textwidth]{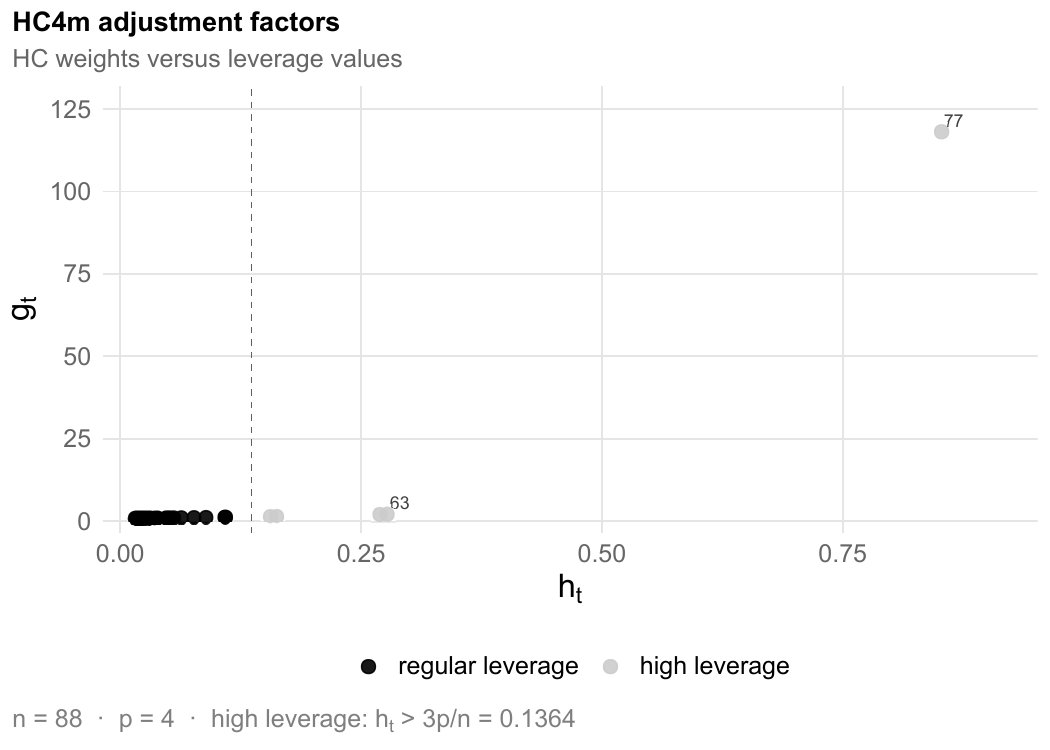}\label{fig:sub_c_02}} &
        \subcaptionbox{\centering}{\includegraphics[width=0.45\textwidth]{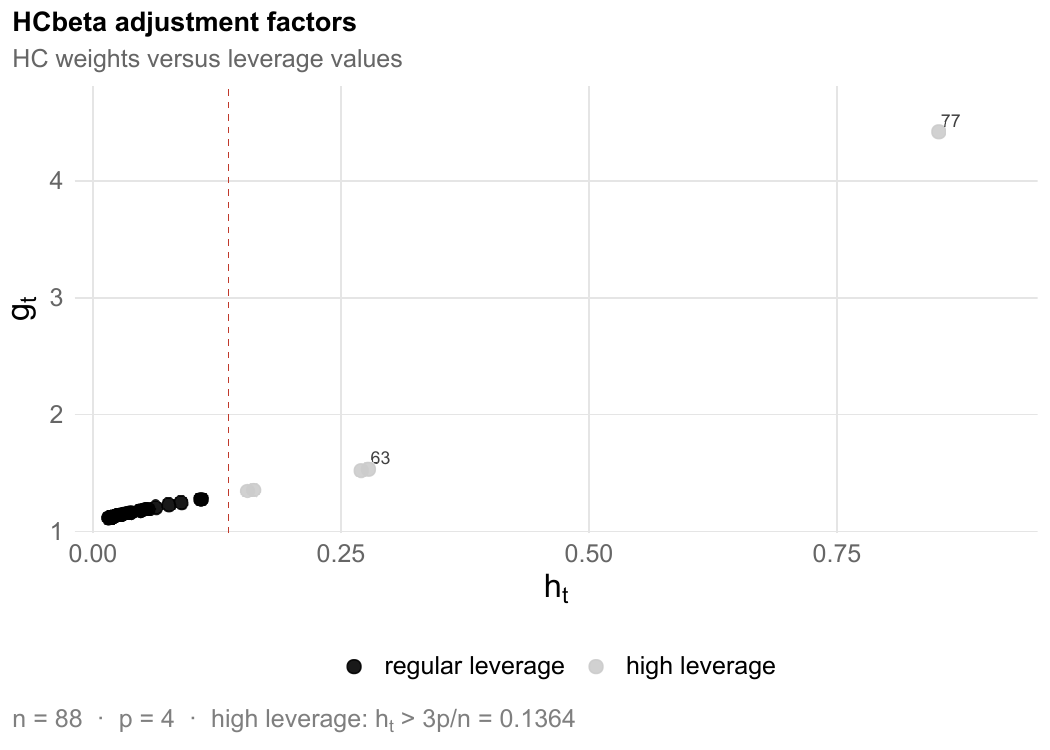}\label{fig:sub_d_02}}
    \end{tabular}

    \caption{Adjustment factors ($g_t$) versus leverages ($h_t$) for HC3 (top-left panel, A), HC4 (top-right panel, B), HC4m (bottom-left panel, C), and $\text{HC}_\beta$ (bottom-right panel, D), complete data; house prices dataset.}
    \label{F:leverages_02}
\end{figure}

Figure~\ref{F:leverages_02} further illustrates these findings by plotting the adjustment factors ($g_t$) against leverage ($h_t$) for each estimator. The overshooting problem is clearly visible for HC3 (top-left panel), HC4 (top-right panel), and HC4m (bottom-left panel), whose maximum $g_t$ values reach 45.4824, 2{,}068.6510, and 118.1148, respectively. These extreme values stem from the inclusion of observations $63$ and $77$, both exhibiting high leverage. In contrast, the adjustment factor corresponding to the $\text{HC}_\beta$ estimator (bottom-right panel) displays a much smoother and controlled growth, with a maximum $g_t$ of only 4.4184, demonstrating superior stability in the presence of influential leverage points.

\subsection{2009 crime data}

In this application, we analyze U.S.\ crime data for the year 2009. The dataset contains information for the 50 U.S. states and the District of Columbia, totaling 51 observations. The response variable ($y$) is the rate of murders per 100{,}000 persons in the population (\texttt{murder}). The explanatory variables are: the percentage of the population having graduated from high school or higher (\texttt{hs\_grad}), the percentage of individuals living below the poverty line (\texttt{poverty}), and the percentage of families composed of single individuals (\texttt{single}). The regression model fitted to the data is
\begin{equation*}
y_t = \beta_1 + \beta_2 x_{t2} + \beta_3 x_{t3} + \beta_4 x_{t4} + e_t,
\quad t = 1, \ldots, 51,
\end{equation*}
where $x_{2}$ corresponds to \texttt{hs\_grad}, $x_{3}$ to \texttt{poverty}, and $x_{4}$ to \texttt{single}. The OLS estimates of the regression coefficients are $\hat{\beta}_1 = -40.6531$, $\hat{\beta}_2 = 0.2755$, $\hat{\beta}_3 = 0.3530$, and $\hat{\beta}_4 = 0.6642$. The fitted model yields a coefficient of determination of $R^2 = 0.8112$. The presence of heteroskedasticity is confirmed by the Breusch--Pagan test ($p\text{-value} = 0.0163$), highlighting the need for heteroskedasticity-consistent standard errors.

The sample contains a single observation whose leverage exceeds the threshold $3p/n = 0.2353$: the District of Columbia ($h_9 = 0.7365$). This extremely high leverage value indicates that this observation exerts disproportionate influence on the fitted model. For this reason, we also examine results obtained after removing this observation from the sample.

The OLS, HC0, HC3, HC4, HC4m, bootstrap, and $\text{HC}_\beta$ standard errors of $\hat{\beta}_1$, $\hat{\beta}_2$, $\hat{\beta}_3$, and $\hat{\beta}_4$ are reported in Table~\ref{tab:se_empirico_crime}. The presence of the District of Columbia, with leverage close to $0.74$, causes a dramatic inflation of the HC4 standard errors, which reach (in the complete sample) $133.26$ for $\hat{\beta}_1$ and $1.1678$ for $\hat{\beta}_2$. The HC4m estimator also produces very large values ($49.44$ and $0.4344$, respectively), although to a lesser extent. In contrast, the $\text{HC}_\beta$ estimator yields considerably more stable standard errors ($25.39$ and $0.2253$), which are close to the pairs bootstrap values ($20.91$ and $0.1866$), containing the impact of the influential observation while avoiding excessive corrections. The fitted Beta distribution yields parameter estimates $\tilde{a} = 3.4252$ and $\tilde{b} = 0.7444$.

Notably, the $\text{HC}_\beta$ standard errors computed from the complete and incomplete samples exhibit greater agreement than those obtained from HC3, HC4, and HC4m. For example, when the leverage point is removed, the $\text{HC}_\beta$ standard error of $\hat{\beta}_2$ decreases from $0.2253$ to $0.1616$. By contrast, the HC3, HC4, and HC4m standard errors change from $0.3150$, $1.1678$, and $0.4344$ to $0.1352$, $0.1369$, and $0.1398$, respectively. An even more striking pattern emerges for $\hat{\beta}_1$: the $\text{HC}_\beta$ standard error decreases from $25.3926$ to $18.6469$, whereas the HC3, HC4, and HC4m standard errors fall from $35.7440$, $133.2648$, and $49.4370$ to $15.4584$, $15.7612$, and $16.0113$, respectively. These large swings highlight the sensitivity of conventional leverage-adjusted HC estimators to extreme leverage, whereas the $\text{HC}_\beta$ estimator displays a noticeably more stable behavior.

\begin{table}[htb]
    \centering
    \begin{threeparttable}
    \caption{Standard errors with complete data and without the leverage point; 2009 crime dataset.}
    \label{tab:se_empirico_crime}
    \begin{tabular}{@{}l *{7}{r}@{}}
        \toprule
        & \multicolumn{1}{c}{OLS} & \multicolumn{1}{c}{HC0}
        & \multicolumn{1}{c}{HC3} & \multicolumn{1}{c}{HC4}
        & \multicolumn{1}{c}{HC4m} & \multicolumn{1}{c}{$\text{HC}_\beta$}
        & \multicolumn{1}{c}{$\text{boot}_p$} \\
        \midrule
        \multicolumn{7}{l}{\textit{Complete sample}} \\[0.3em]
        $\text{se}(\hat{\beta}_1)$ & 10.9992 & 11.2029 &  35.7440 & 133.2648 & 49.4379 & 25.3926 & 20.9092 \\
        $\text{se}(\hat{\beta}_2)$ &  0.1084 &  0.1034 &   0.3150 &   1.1678 &  0.4344 &  0.2253 & 0.1866 \\
        $\text{se}(\hat{\beta}_3)$ &  0.1120 &  0.1158 &   0.1830 &   0.5231 &  0.2263 &  0.1589 & 0.1350 \\
        $\text{se}(\hat{\beta}_4)$ &  0.0589 &  0.0802 &   0.2696 &   1.0116 &  0.3741 &  0.1900 & 0.1551 \\[0.6em]
        \multicolumn{7}{l}{\textit{Incomplete sample}} \\[0.3em]
        $\text{se}(\hat{\beta}_1)$ & 14.7431 & 13.2017 & 15.4584 & 15.7612 & 16.0113 & 18.6469 & 14.7440 \\
        $\text{se}(\hat{\beta}_2)$ &  0.1369 &  0.1161 &  0.1352 &  0.1369 &  0.1398 &  0.1616 & 0.1309 \\
        $\text{se}(\hat{\beta}_3)$ &  0.1101 &  0.1112 &  0.1318 &  0.1365 &  0.1370 &  0.1624 & 0.1185 \\
        $\text{se}(\hat{\beta}_4)$ &  0.0981 &  0.1125 &  0.1295 &  0.1298 &  0.1333 &  0.1520 & 0.1202 \\
        \bottomrule
    \end{tabular}
    \end{threeparttable}
\end{table}

\begin{figure}[htb]
    \centering
    \begin{tabular}{cc}
        \subcaptionbox{\centering}{\includegraphics[width=0.45\textwidth]{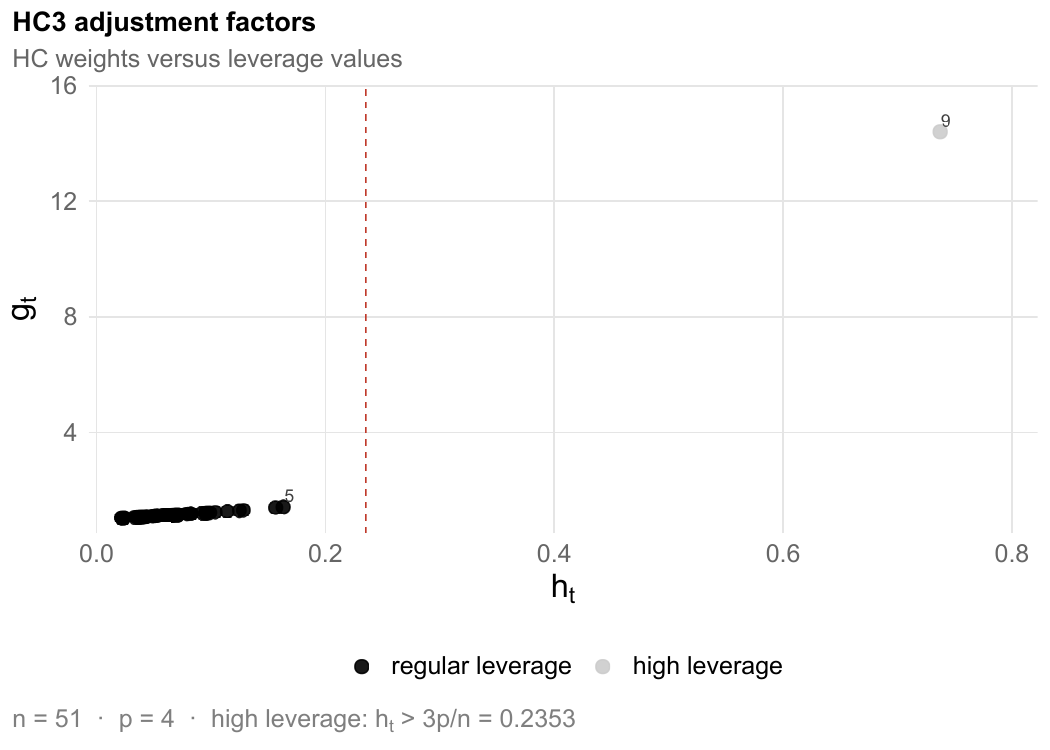}\label{fig:sub_a_04}} &
        \subcaptionbox{\centering}{\includegraphics[width=0.45\textwidth]{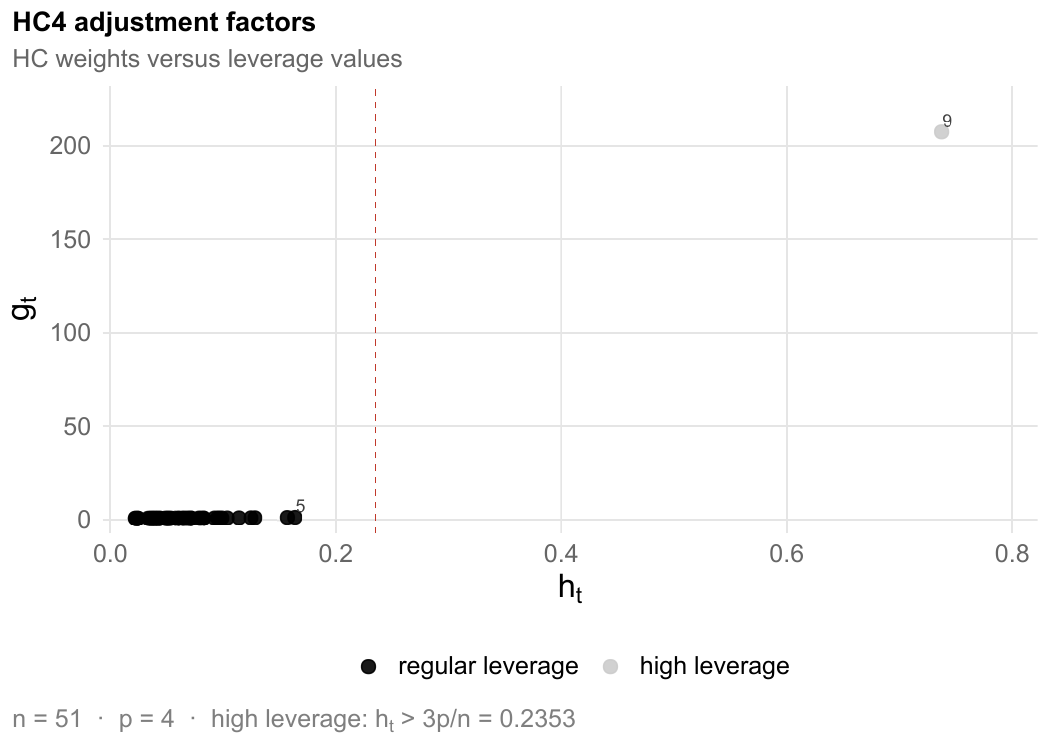}\label{fig:sub_b_04}} \\[1em]

        \subcaptionbox{\centering}{\includegraphics[width=0.45\textwidth]{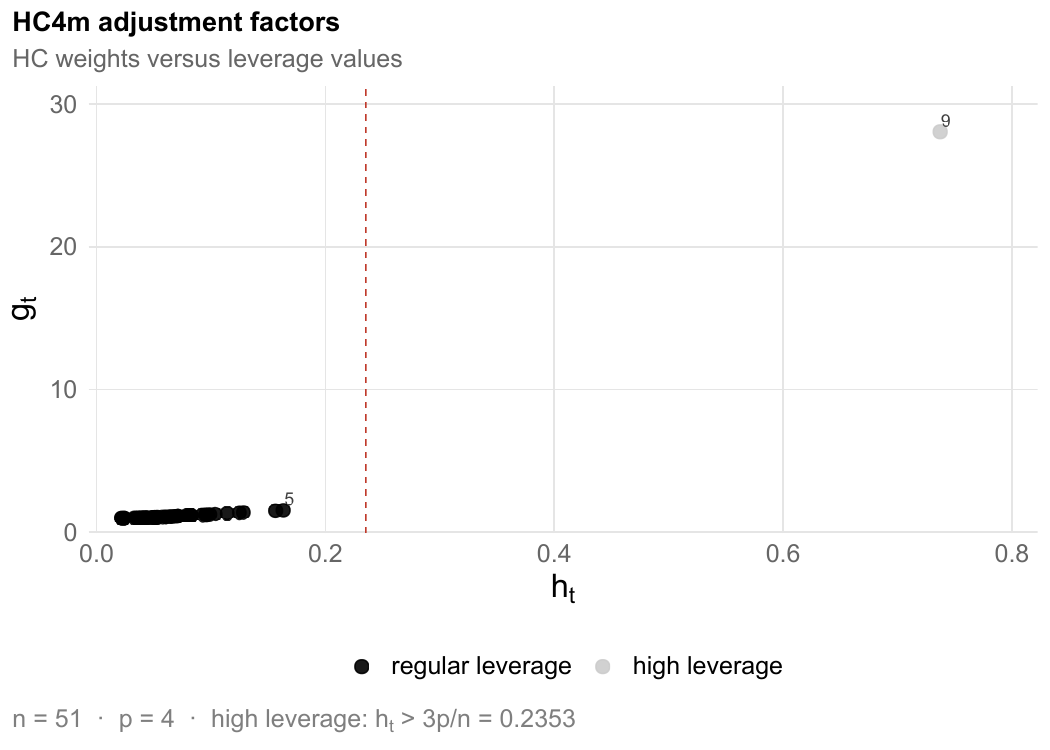}\label{fig:sub_c_04}} &
        \subcaptionbox{\centering}{\includegraphics[width=0.45\textwidth]{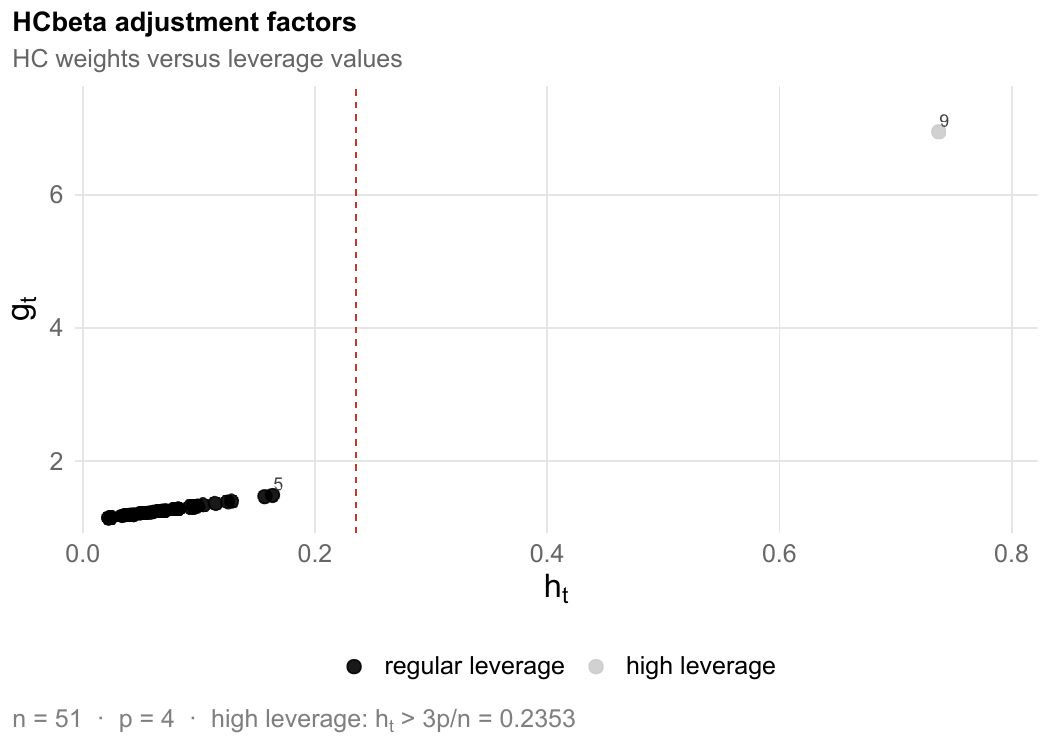}\label{fig:sub_d_04}}
    \end{tabular}
    \caption{Adjustment factors ($g_t$) versus leverages ($h_t$) for HC3 (top-left panel, A), HC4 (top-right panel, B), HC4m (bottom-left panel, C), and $\text{HC}_\beta$ (bottom-right panel, D), complete data; 2009 crime dataset.}
    \label{F:leverages_04}
\end{figure}

A clearer perspective emerges from Figure~\ref{F:leverages_04}, which plots the adjustment factors ($g_t$ versus $h_t$) for the HC3, HC4, HC4m, and $\text{HC}_\beta$ estimators, cases 5 and 9 corresponding to California and the District of Columbia, respectively. The contrast across panels is striking. HC4 (panel B) reaches values of $g_t$ exceeding $200$ for the District of Columbia, while HC4m (panel C) exceeds $25$. Even HC3 (panel A), though less extreme, attains values close to $14$. In contrast, the adjustment factor associated with $\text{HC}_\beta$ (panel D) remains below $7$ for the same observation. This behavior illustrates the ability of $\text{HC}_\beta$ to accommodate extremely high leverage points without disproportionately amplifying the corrections applied to squared residuals and, consequently, without undermining inferential stability.

The impact of the District of Columbia on statistical inference becomes even clearer when examining the $p$-values for the test of $\mathcal{H}_0\colon \beta_4 = 0$ against $\mathcal{H}_1\colon \beta_4 \neq 0$, reported in Table~\ref{tab:pvalues_beta4_crime}. In the complete sample, the HC4 and HC4m tests fail to reject $\mathcal{H}_0$, yielding $p$-values of $0.512$ and $0.076$, respectively. In contrast, the $\text{HC}_\beta$ test rejects $\mathcal{H}_0$ at the $1\%$ significance level ($p\text{-value} < 0.001$), aligning with the inference obtained after removing the influential observation from the dataset. Once the District of Columbia is excluded, all estimators produce similar standard errors and all tests reject $\mathcal{H}_0 \colon \beta_4 = 0$ at the $1\%$ significance level. Overall, the $\text{HC}_\beta$-based test appears to provide the most coherent inference when the full dataset is used, as the estimator avoids the excessive inflation of adjustment factors that can arise in the presence of extreme leverage.

\begin{table}[htb]
\centering
\begin{threeparttable}
\caption{$p$-values for the test $\mathcal{H}_0\colon \beta_4 = 0$; 2009 crime dataset.}
\label{tab:pvalues_beta4_crime}
\begin{tabular}{@{}l r l r@{}}
    \toprule
    \multicolumn{2}{c}{Complete data, $n = 51$} &
    \multicolumn{2}{c}{Incomplete data, $n = 50$} \\
    \cmidrule(lr){1-2} \cmidrule(lr){3-4}
    Test & $p$-value & Test & $p$-value \\
    \midrule
    HC0          & $<$0.0001 & HC0          & 0.0004 \\
    HC3          & 0.0138 & HC3          & 0.0021 \\
    HC4          & 0.5115 & HC4          & 0.0022 \\
    HC4m         & 0.0758 & HC4m         & 0.0028 \\
    $\text{HC}_\beta$ & 0.0005 & $\text{HC}_\beta$ & 0.0088 \\
    $\text{boot}_p$ & $<$0.0001 & $\text{boot}_p$ & 0.0009 \\
    \bottomrule
\end{tabular}
\end{threeparttable}
\end{table}

\subsection{Orthorexia nervosa and exercise addiction}

It is instructive to examine the performance of the estimator proposed in this article in a setting where no observations exhibit extreme leverage, although several cases may still be regarded as moderately high-leverage points. For this purpose, we consider the dataset analyzed by \citet{Wachten+Strahler_2026}, who investigated the relationship between orthorexia nervosa (OrNe) and exercise addiction (EA) in a sample of $n = 384$ participants. The dependent variable ($y$) measures psychosocial impairment through the total score of the CIA (\textit{Clinical Impairment Assessment}), an instrument designed to evaluate the extent to which eating and exercise habits have affected the participant's life during the four weeks preceding the survey. The explanatory variables include: the standardized score of the orthorexia nervosa subscale of the \textit{Teruel Orthorexia Scale} (TOS-OrNe); the standardized score of restrictive eating behaviors (EDE-Q); the standardized score of muscularity-oriented behaviors (MOET); and gender, included as a control variable.  The regression model considered is
\begin{equation*}
y_t = \beta_1 + \beta_2 x_{t2} + \beta_3 x_{t3} + \beta_4 x_{t4} + \beta_5 x_{t5}
+ e_t, \quad t = 1, \ldots, 384,
\end{equation*}
where $x_{2}$ denotes the EDE-Q score, $x_{3}$ the MOET score, $x_{4}$ the TOS-OrNe score, and $x_{5}$ the participant's gender. The OLS estimates of the regression coefficients are $\hat{\beta}_1 = -0.0456$, $\hat{\beta}_2 = 0.5459$, $\hat{\beta}_3 = 0.1472$, $\hat{\beta}_4 = 0.2114$, and $\hat{\beta}_5 = 0.0598$. There are 21 observations with leverage values exceeding the threshold $3p/n = 0.0391$, among which observation 132 stands out with $h_{132} = 0.0962$. The fitted model yields a coefficient of determination of $R^2 = 0.6593$. The presence of heteroskedasticity is confirmed by the Breusch--Pagan test ($p\text{-value} = 3.239 \times 10^{-7}$).

The estimated parameters $\tilde{a} = 101.7362$ and $\tilde{b} = 1.6736$ differ substantially from the uniform benchmark ($a = b = 1$), as well as from the estimates obtained in previous applications, where the presence of highly influential observations was typically associated with smaller values of $\tilde{a}$. In the present dataset, the large value of $\tilde{a}$ is consistent with a configuration in which leverage values are uniformly small and show limited dispersion across observations. In this setting, the complementary leverages are concentrated near one, and the fitted Beta distribution provides a flexible description of this concentration without imposing additional structural assumptions.

The OLS, HC0, HC3, HC4, HC4m, bootstrap, and $\text{HC}_\beta$ standard errors of $\hat{\beta}_1, \ldots, \hat{\beta}_5$ are reported in Table~\ref{tab:se_empirico_02}. In this dataset, characterized by moderate leverage and the absence of extremely influential observations, $\text{HC}_\beta$ produces the largest standard errors among the estimators considered. For instance, the standard error of $\hat{\beta}_3$ equals $0.0700$ (HC0), $0.0731$ (HC3), $0.0754$ (HC4), $0.0738$ (HC4m), $0.0711$ (pairs bootstrap), and $0.0806$ ($\text{HC}_\beta$). This pattern reflects the adaptive mechanism underlying the estimator: when leverage values remain within a moderate range, the moment-matched Beta distribution induces a somewhat stronger correction, leading to slightly larger standard errors. In this way, $\text{HC}_\beta$ avoids the risk of underestimating sampling variability while still maintaining stable adjustments across observations.

\begin{table}[htb]
    \centering
    \begin{threeparttable}
    \caption{Standard errors with complete data; orthorexia nervosa and exercise addiction dataset.}
    \label{tab:se_empirico_02}
    \begin{tabular}{@{}l *{7}{r}@{}}
        \toprule
        & \multicolumn{1}{c}{OLS} & \multicolumn{1}{c}{HC0} & \multicolumn{1}{c}{HC3}
        & \multicolumn{1}{c}{HC4} & \multicolumn{1}{c}{HC4m}
        & \multicolumn{1}{c}{$\text{HC}_\beta$} & \multicolumn{1}{c}{$\text{boot}_p$}\\
        \midrule
        $\text{se}(\hat{\beta}_1)$ & 0.0638 & 0.0541 & 0.0560 & 0.0572 & 0.0564 & 0.0606 & 0.0543 \\
        $\text{se}(\hat{\beta}_2)$ & 0.0415 & 0.0689 & 0.0718 & 0.0739 & 0.0726 & 0.0792 & 0.0687 \\
        $\text{se}(\hat{\beta}_3)$ & 0.0471 & 0.0700 & 0.0731 & 0.0754 & 0.0738 & 0.0806 & 0.0711 \\
        $\text{se}(\hat{\beta}_4)$ & 0.0471 & 0.0687 & 0.0718 & 0.0742 & 0.0725 & 0.0793 & 0.0696 \\
        $\text{se}(\hat{\beta}_5)$ & 0.0738 & 0.0679 & 0.0701 & 0.0715 & 0.0706 & 0.0758 & 0.0678 \\
        \bottomrule
    \end{tabular}
    \end{threeparttable}
\end{table}

The main inferential interest lies in testing $\mathcal{H}_0\colon \beta_3 = 0$ against $\mathcal{H}_1\colon \beta_3 \neq 0$, that is, assessing whether muscularity-oriented behaviors exert a significant effect on psychosocial impairment. Using the full sample, the conventional $z$ test (based on the OLS standard error) yields a $p$-value of $0.0018$, leading to rejection of the null hypothesis at the $1\%$ significance level.

With the complete sample, the tests based on HC0, HC3, and HC4m reject the null hypothesis at the $5\%$ level, with $p$-values of $0.0355$, $0.0440$, and $0.0462$, respectively. The quasi-$t$ test based on the pairs bootstrap also yields rejection of the null hypothesis at the $5\%$ significance level ($p\text{-value} = 0.0384$). In contrast, the HC4 and $\text{HC}_\beta$ tests do not reject $\mathcal{H}_0$, yielding $p$-values of $0.0508$ and $0.0678$. Although both procedures lead to the same formal decision, the HC4 result lies very close to the conventional $5\%$ significance threshold, whereas the $\text{HC}_\beta$ test provides stronger evidence in favor of non-rejection.

In the reduced sample, obtained by removing the 21 observations with $h_t > 3p/n$, the $p$-values for testing $\mathcal{H}_0\colon \beta_3 = 0$ become $0.0684$ (OLS), $0.1318$ (HC0), $0.1453$ (HC3), $0.1527$ (HC4), $0.1484$ (HC4m), $0.1322$ (pairs bootstrap), and $0.1901$ ($\text{HC}_\beta$). Importantly, the $\text{HC}_\beta$ test yields the same inferential conclusion in both the complete and reduced samples, consistently failing to reject $\mathcal{H}_0$. The corresponding $p$-values remain relatively stable despite the removal of all leverage points identified by the $3p/n$ criterion. HC4 also leads to non-rejection in both samples; however, its complete-sample result is much closer to the rejection boundary. Overall, these findings suggest that inference based on $\text{HC}_\beta$ is comparatively less affected by moderate changes in the leverage structure of the data, thereby providing greater inferential stability.

\begin{figure}[htb]
    \centering
    \begin{tabular}{cc}
        \subcaptionbox{\centering}{\includegraphics[width=0.45\textwidth]{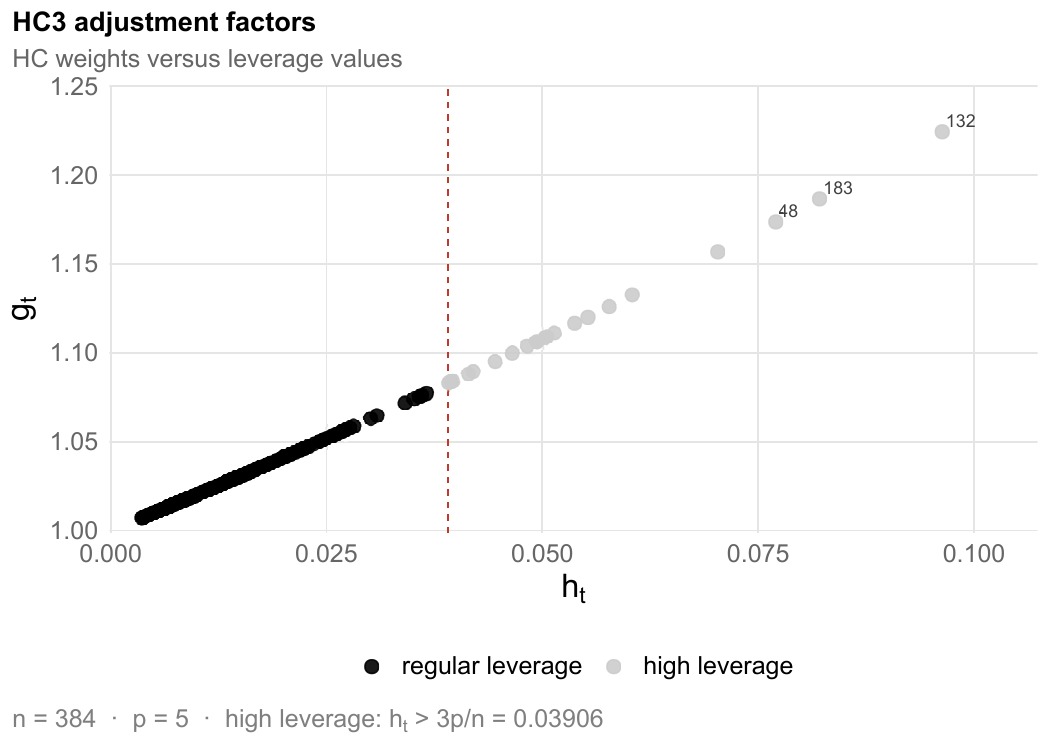}\label{fig:sub_a_03}} &
        \subcaptionbox{\centering}{\includegraphics[width=0.45\textwidth]{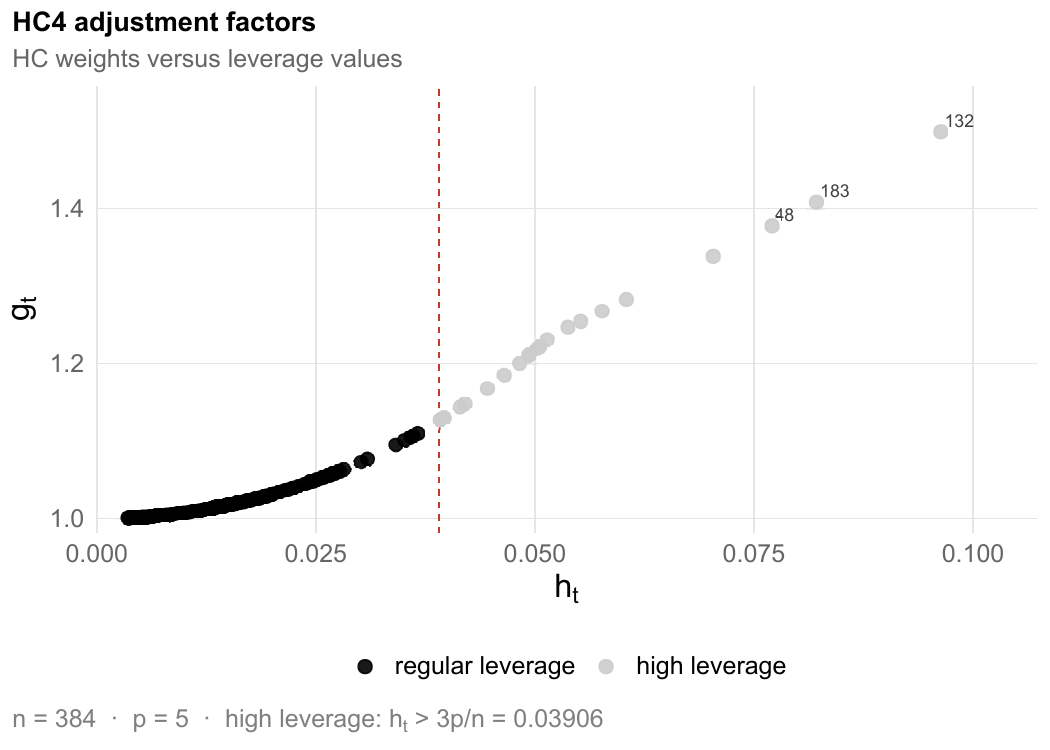}\label{fig:sub_b_03}} \\[1em]

        \subcaptionbox{\centering}{\includegraphics[width=0.45\textwidth]{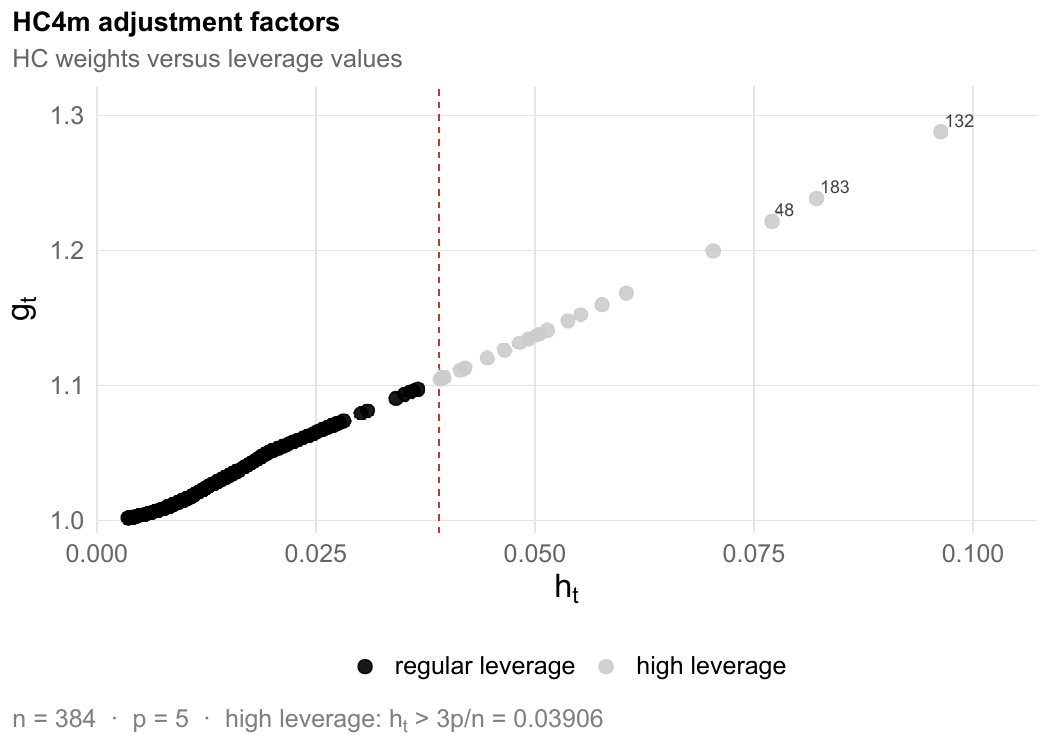}\label{fig:sub_c_03}} &
        \subcaptionbox{\centering}{\includegraphics[width=0.45\textwidth]{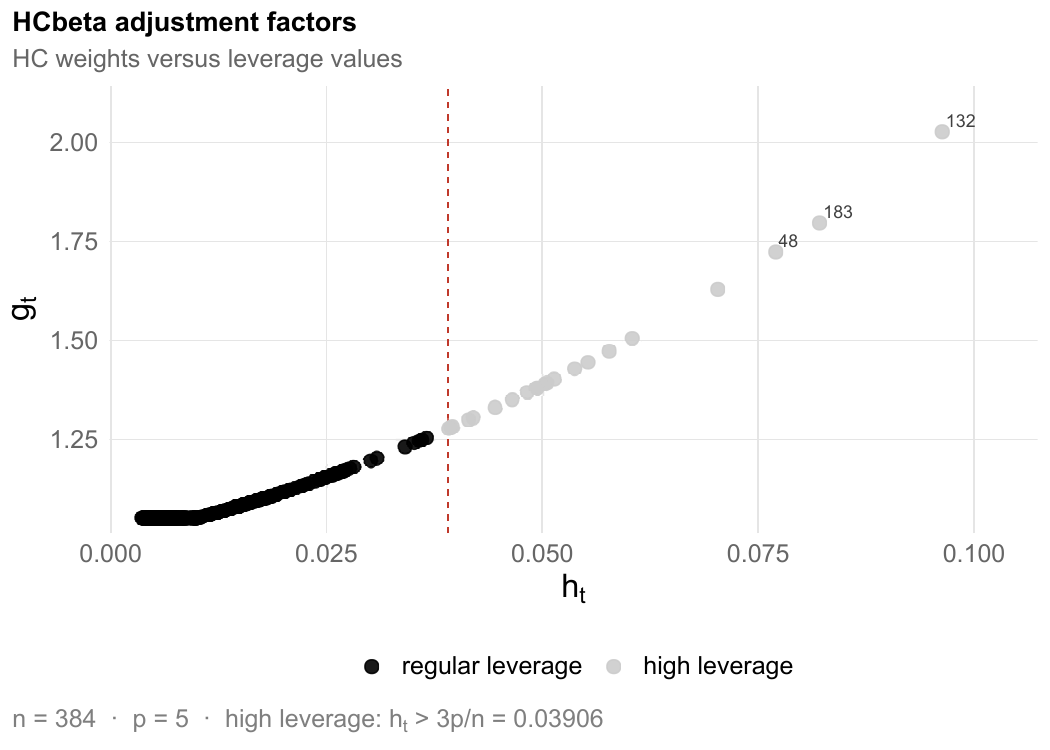}\label{fig:sub_d_03}}
    \end{tabular}
    \caption{Adjustment factors ($g_t$) versus leverages ($h_t$) for HC3 (top-left
    panel, A), HC4 (top-right panel, B), HC4m (bottom-left panel, C), and
    $\text{HC}_\beta$ (bottom-right panel, D), complete data; orthorexia nervosa and
    exercise addiction dataset.}
    \label{F:leverages_03}
\end{figure}

Figure~\ref{F:leverages_03} displays the adjustment factors ($g_t$ versus $h_t$) for the HC3 (top-left panel), HC4 (top-right panel), HC4m (bottom-left panel), and $\text{HC}_\beta$ (bottom-right panel) estimators. Unlike what is observed in the previous applications, the absence of extreme leverage points in this dataset --- with $h_t$ values below $0.10$ --- results in relatively moderate adjustment factors for all estimators. Nevertheless, the growth of $g_t$ as a function of $h_t$ is smoother and more gradual for $\text{HC}_\beta$ than for the alternative estimators, with maximum values of $2.0268$ ($\text{HC}_\beta$), $1.2243$ (HC3), $1.4989$ (HC4), and $1.2878$ (HC4m). In samples characterized by moderate leverage, the estimator applies corrections in a controlled manner, preserving inferential stability without inducing abrupt changes in the adjustment factors.

This behavior illustrates a key property of the $\text{HC}_\beta$ estimator: by governing the severity of the corrections through the Beta cumulative distribution function, the estimator adapts smoothly to the leverage structure of the sample. As a result, it often preserves the coherence of statistical inference even when the sample composition changes moderately. In contrast, HC3, HC4, and HC4m rely on less flexible adjustment mechanisms, which can make them more sensitive to variations in the leverage configuration of the data.

\section{Computer implementation}\label{S:r-package}

We developed the \textsl{hcinfer} package for \textsf{R}, which implements the proposed $\text{HC}_\beta$ estimator as well as several alternative heteroskedasticity-consistent covariance matrix estimators. The package is publicly available through the \textit{Comprehensive R Archive Network} (CRAN) and can be installed directly via \texttt{install.packages("hcinfer")} (see \url{https://CRAN.R-project.org/package=hcinfer}). The development version can be installed using \texttt{remotes::install\_github("prdm0/hcinfer", force = TRUE)} through the \textsl{remotes} package. Comprehensive documentation, including vignettes with reproducible examples illustrating the available estimators and inferential procedures, is available at \url{https://prdm0.github.io/hcinfer/}.

The package provides functions for computing heteroskedasticity-consistent covariance matrix estimators for $\bm{\hat{\beta}}$, constructing confidence intervals for regression coefficients, performing hypothesis tests, generating diagnostic plots, and carrying out related inferential analyses. The typical workflow is straightforward:
\begin{verbatim}
fit <- lm(y ~ x1 + x2, data = data)
result <- hcinfer(fit, type = "hcbeta")
summary(result)
tests(result)
confint(result)
plot(result)

cov <- vcov_hc(fit, type = "hcbeta")
plot(cov)
\end{verbatim}

It is worth noting that $\text{HC}_\beta$ is the default estimator in the package. Consequently, calling \texttt{hcinfer(fit)} without explicitly specifying the \texttt{type} argument automatically invokes \texttt{type = "hcbeta"}. The constants $c_1$ and $c_2$, which appear in the exponent $c_1/n^{c_2}$ of the adjustment factor $g_t$ (Section~\ref{S:hc-beta}), can be customized through the arguments \texttt{c1} and \texttt{c2}. Based on extensive Monte Carlo evidence, we recommend the default values $c_1 = 7$ and $c_2 = 0.75$, which provide a favorable balance between finite-sample performance and asymptotic behavior. Nevertheless, users are free to modify these values to suit their specific applications or methodological preferences.

The \verb|hcinfer()| function accepts any object of class \verb|lm|. In addition to the proposed $\text{HC}_\beta$ estimator, the package implements eight alternative heteroskedasticity-consistent covariance matrix estimators through the \verb|type| argument, namely HC0, HC1, HC2, HC3, HC4, HC4m, HC5, and HC5m. The available estimators can be listed at any time using \verb|hc_methods()|.

The \verb|summary()| function produces a comprehensive report that includes quasi-$t$ tests for the regression coefficients, heteroskedasticity-consistent confidence intervals, and, for each observation, the leverage values $h_t$ together with the adjustment factors $g_t$ associated with the selected estimator. These quantities provide useful diagnostic information and facilitate the identification of potentially influential observations.

The \verb|tests()| function returns only the quasi-$t$ test results in tibble format, offering a direct and convenient way to extract inferential summaries. Individual coefficients can be selected through the \verb|parm| argument, and the significance level can be adjusted via \verb|alpha|. Similarly, \verb|confint()| extracts heteroskedasticity-consistent confidence intervals, supporting the same \verb|parm| argument, while the confidence level is controlled through \verb|level|.

The \verb|vcov_hc()| function provides an alternative workflow when only the heteroskedasti\-city-consistent covariance matrix estimate is required, without performing full inferential procedures. Diagnostic graphics are available through the \verb|plot()| method, which displays robust confidence intervals for objects of class \verb|hcinfer| and leverage values together with HC adjustment factors for objects of class \verb|vcov_hc|.

It is also worth emphasizing that \verb|plot()| returns an object of class \verb|ggplot|. Consequently, the resulting graphics can be readily customized using the extensive functionality available in the \textsl{ggplot2} ecosystem. The adjustment-factor plots ($g_t$ versus $h_t$) presented in the empirical applications of Section~\ref{S:applications} (Figures~\ref{F:leverages_01}, \ref{F:leverages_02}, \ref{F:leverages_04}, and~\ref{F:leverages_03}) were generated using the command \verb|plot(vcov_hc(fit, type = ...))|, where \verb|fit| denotes the fitted \verb|lm| object for the corresponding dataset.

In addition, the \textsl{hcinfer} package provides bootstrap-based standard error estimation and confidence interval construction via the pairs bootstrap, with optional support for parallel computing.

\section{Concluding remarks}\label{S:conclusions}

This paper introduced $\text{HC}_\beta$, a new heteroskedasticity-consistent covariance matrix estimator for ordinary least squares regression. The proposed estimator combines the familiar sandwich structure of traditional HC estimators with a new leverage-adjustment mechanism based on the Beta distribution. Unlike existing approaches, whose corrections are typically driven by fixed or piecewise functions of $(1-h_t)$, the proposed method incorporates a flexible data-driven adjustment that adapts to the empirical distribution of leverage values in the sample. This construction allows the estimator to accommodate heterogeneous leverage patterns while avoiding excessively aggressive corrections.

The Monte Carlo simulation results indicate that the quasi-$t$ tests and confidence intervals based on $\text{HC}_\beta$ display very favorable finite-sample behavior. In particular, the proposed estimator generally delivers smaller size distortions and more accurate coverage rates than competing HC estimators, especially in settings involving strong heteroskedasticity, moderate sample sizes, and high-leverage observations. The simulations also suggest that the inferential procedures based on $\text{HC}_\beta$ converge relatively quickly toward their nominal asymptotic properties as the sample size increases. At the same time, the estimator tends to remain mildly conservative in some scenarios, providing additional protection against spurious rejections.

The empirical applications further complement the Monte Carlo evidence by illustrating the practical relevance of the proposed methodology across datasets with markedly different leverage structures. In datasets containing highly influential observations, traditional leverage-adjusted estimators such as HC3, HC4, and HC4m occasionally produced extremely large adjustment factors (overshooting) and inflated standard errors, leading to unstable inferential conclusions. In contrast, $\text{HC}_\beta$ consistently produced more stable standard errors and more coherent inferential results across both complete and incomplete samples. The applications also highlight an important feature of the proposed estimator: its ability to adapt smoothly to different leverage configurations without requiring ad hoc tuning rules or preliminary decisions regarding the severity of leverage corrections. Although these empirical examples are intended to illustrate rather than establish finite-sample superiority, they reinforce the simulation results by demonstrating that the adaptive construction of $\text{HC}_\beta$ translates into stable and interpretable inference in realistic regression settings.

Taken together, the simulation study and the empirical applications provide complementary evidence of the practical value of the proposed estimator: the former quantifies its finite-sample inferential properties under controlled conditions, whereas the latter demonstrates its behavior in realistic data analysis problems.

From a broader perspective, the results obtained in this paper suggest that excessive inflation of residual-based corrections under extreme leverage may constitute an important source of inferential instability in heteroskedasticity-consistent procedures. The proposed estimator addresses this issue by introducing a smoother and more adaptive weighting mechanism. In this sense, $\text{HC}_\beta$ may be viewed not merely as another member of the HC family, but as an alternative framework for constructing leverage-sensitive covariance matrix estimators.

From a practical perspective, $\text{HC}_\beta$ may be viewed as a data-adaptive default among heteroskedasticity-consistent covariance matrix estimators. When leverage values are relatively homogeneous, it performs similarly to existing HC estimators. Its main practical advantage emerges in regression designs exhibiting heterogeneous or unusually large leverage values, where fixed correction schemes may produce either insufficient adjustment or excessive growth of the adjustment factors. In such situations, HC$_\beta$ provides a more stable compromise between finite-sample accuracy and protection against overly aggressive corrections.

Unless there is a specific reason to reproduce a traditional HC estimator, we recommend HC$_\beta$ as the default option for applied analyses because it automatically adapts its adjustment to the observed leverage configuration. The empirical comparisons with the pairs bootstrap reported in this paper provide additional evidence that this adaptive strategy yields practically relevant improvements while avoiding the excessive adjustment factors often produced by more aggressive HC estimators.

The present paper should be viewed as introducing a general
distribution-based approach to leverage adjustment, with the Beta family
serving as a natural and computationally convenient starting point rather than
as the only possible choice. Exploring alternative distributions on the unit
interval, such as the Kumaraswamy family, or different estimation strategies
such as L-moments, constitutes an interesting avenue for future research.

To facilitate practical use of the methodology, the proposed estimator and related inferential tools were implemented in the publicly available \textsf{R} package \textsl{hcinfer}. The package provides a unified computational framework for estimation, hypothesis testing, confidence interval construction, and graphical diagnostics, making the proposed procedures readily accessible to applied researchers.

\section*{Acknowledgments}

The authors are grateful to Hanna Wachten and Jana Strahler for kindly providing the dataset used in the fourth empirical application. They also thank Klaus Vasconcellos for helpful discussions on the topic of this paper. Finally, the authors are grateful to the two anonymous reviewers for their careful reading of the manuscript and for their constructive comments and suggestions, which substantially improved the quality and clarity of the paper.

\section*{Disclosure statement}

No potential conflict of interest was reported by the authors.

\section*{Funding}

Marina O.~Cunha's research was funded by Coordenação de Aperfeiçoamento de Pessoal de Nível Superior (Finance Code 001) and Francisco Cribari-Neto's research was funded by Conselho Nacional de Desenvolvimento Científico e Tecnológico (grant
304646/2023-7).

\section*{Author contributions}

CRediT: \textbf{Marina O.~Cunha}: Methodology, Formal analysis, Data Curation, Software, Validation, Investigation, Writing - Original Draft, Writing - Review \& Editing, Visualization; \textbf{Francisco Cribari-Neto}: Conceptualization, Methodology, Formal analysis, Software, Validation, Investigation, Resources, Writing - Original Draft, Writing - Review \& Editing, Visualization, Supervision, Project administration, Funding acquisition; \textbf{Pedro~D.~R.~Marinho:} Data Curation, Software, Validation, Investigation, Visualization.


\begin{thebibliography}{21}
\providecommand{\natexlab}[1]{#1}
\providecommand{\url}[1]{\texttt{#1}}
\expandafter\ifx\csname urlstyle\endcsname\relax
  \providecommand{\doi}[1]{doi: #1}\else
  \providecommand{\doi}{doi: \begingroup \urlstyle{rm}\Url}\fi

\bibitem[Chesher and Jewitt(1987)]{Chesher+Jewitt_1987}
A.~Chesher and I.~Jewitt.
\newblock The bias of a heteroskedasticity consistent covariance matrix
  estimator.
\newblock \emph{Econometrica}, 55\penalty0 (5):\penalty0 1217--1222, 1987.
\newblock \doi{10.2307/1911269}.

\bibitem[Cribari-Neto(2004)]{Cribari_2004}
F.~Cribari-Neto.
\newblock Asymptotic inference under heteroskedasticity of unknown form.
\newblock \emph{Computational Statistics \& Data Analysis}, 45\penalty0
  (2):\penalty0 215--233, 2004.
\newblock \doi{10.1016/S0167-9473(02)00366-3}.

\bibitem[Cribari-Neto and Pereira(2019)]{Cribari+Pereira_2019}
F.~Cribari-Neto and I.~F.~S. Pereira.
\newblock Testing inference in heteroskedastic linear regressions: A comparison
  of two alternative approaches.
\newblock \emph{Journal of Statistical Computation and Simulation}, 89\penalty0
  (8):\penalty0 1437--1465, 2019.
\newblock \doi{10.1080/00949655.2019.1586902}.

\bibitem[Cribari-Neto and Silva(2011)]{Cribari+Silva_2011}
F.~Cribari-Neto and W.~B. Silva.
\newblock A new heteroskedasticity-consistent covariance matrix estimator for
  the linear regression model.
\newblock \emph{{AStA} Advances in Statistical Analysis}, 95\penalty0
  (2):\penalty0 129--146, 2011.
\newblock \doi{10.1007/s10182-010-0141-2}.

\bibitem[Cribari-Neto et~al.(2000)Cribari-Neto, Ferrari, and
  Cordeiro]{Cribari+Ferrari+Cordeiro_2000}
F.~Cribari-Neto, S.~L.~P. Ferrari, and G.~M. Cordeiro.
\newblock Improved heteroscedasticity-consistent covariance matrix estimators.
\newblock \emph{Biometrika}, 87\penalty0 (4):\penalty0 907--918, 2000.
\newblock \doi{10.1093/biomet/87.4.907}.

\bibitem[Cribari-Neto et~al.(2007)Cribari-Neto, Souza, and
  Vasconcellos]{Cribari+Souza+Vasconcellos_2007}
F.~Cribari-Neto, T.~C. Souza, and K.~L.~P. Vasconcellos.
\newblock Inference under heteroskedasticity and leveraged data.
\newblock \emph{Communications in Statistics - Theory and Methods}, 36\penalty0
  (10):\penalty0 1877--1888, 2007.
\newblock \doi{10.1080/03610920601126589}.
\newblock Erratum: 37, 2008, 3329-3330.

\bibitem[Davidson and MacKinnon(1993)]{Davidson+MacKinnon_1993}
R.~Davidson and J.~G. MacKinnon.
\newblock \emph{Estimation and Inference in Econometrics}.
\newblock Oxford University Press, New York, 1993.
\newblock ISBN 9780195060119.

\bibitem[Farrar et~al.(2025)Farrar, Blignaut, Luus, and
  Steel]{Farrar+et-al_2025}
T.~Farrar, R.~Blignaut, R.~Luus, and S.~Steel.
\newblock A review and comparison of methods of parameter estimation and
  inference for heteroskedastic linear regression models.
\newblock \emph{Journal of Applied Statistics}, 52\penalty0 (16):\penalty0
  3091--3120, 2025.
\newblock \doi{10.1080/02664763.2025.2496719}.

\bibitem[Flachaire(2005)]{Flachaire_2005}
E.~Flachaire.
\newblock Bootstrapping heteroskedastic regression models: wild bootstrap vs.\
  pairs bootstrap.
\newblock \emph{Computational Statistics \& Data Analysis}, 49\penalty0
  (2):\penalty0 361--376, 2005.
\newblock \doi{10.1016/j.csda.2004.05.018}.

\bibitem[Furno(1996)]{Furno_1996b}
M.~Furno.
\newblock Small sample behavior of a robust heteroskedasticity consistent
  covariance matrix estimator.
\newblock \emph{Journal of Statistical Computation and Simulation}, 54\penalty0
  (1-3):\penalty0 115--128, 1996.
\newblock \doi{10.1080/00949659608811723}.

\bibitem[Hayes and Cai(2007)]{Hayes+Cai_2007}
A.~F. Hayes and L.~Cai.
\newblock Using heteroskedasticity-consistent standard error estimators in
  {OLS} regression: An introduction and software implementation.
\newblock \emph{Behavior Research Methods}, 39\penalty0 (4):\penalty0 709--722,
  2007.
\newblock \doi{10.3758/bf03192961}.

\bibitem[Hinkley(1977)]{Hinkley_1977}
D.~V. Hinkley.
\newblock Jackknifing in unbalanced situations.
\newblock \emph{Technometrics}, 19\penalty0 (3):\penalty0 285--292, 1977.
\newblock \doi{10.1080/00401706.1977.10489550}.

\bibitem[Jones(2009)]{Jones_2009}
M.~C. Jones.
\newblock {K}umaraswamy distribution: A beta-type distribution with some
  tractability advantages.
\newblock \emph{Statistical Methodology}, 6\penalty0 (1):\penalty0 70--81,
  2009.
\newblock \doi{10.1016/j.stamet.2008.04.001}.

\bibitem[Li et~al.(2016)Li, Zhang, Zhang, and Wang]{Li+et-al_2016}
S.~Li, N.~Zhang, X.~Zhang, and G.~Wang.
\newblock A new heteroskedasticity-consistent covariance matrix estimator and
  inference under heteroskedasticity.
\newblock \emph{Journal of Statistical Computation and Simulation}, 87\penalty0
  (1):\penalty0 198--210, 2016.
\newblock \doi{10.1080/00949655.2016.1198906}.

\bibitem[MacKinnon and White(1985)]{MacKinnon+White_1985}
J.~G. MacKinnon and H.~White.
\newblock Some heteroskedasticity-consistent covariance matrix estimators with
  improved finite-sample properties.
\newblock \emph{Journal of Econometrics}, 29\penalty0 (3):\penalty0 305--325,
  1985.
\newblock \doi{10.1016/0304-4076(85)90158-7}.

\bibitem[Marinho et~al.(2025)Marinho, Cribari-Neto, and
  Tomazella]{Marinho+Cribari+Tomazella_2025}
P.~R.~D. Marinho, F.~Cribari-Neto, and V.~Tomazella.
\newblock Theory and computational tool for interval estimation in linear
  regressions under heteroscedasticity of unknown form using double bootstrap
  methods.
\newblock \emph{Communications in Statistics - Theory and Methods}, 54\penalty0
  (24):\penalty0 7986--8013, 2025.
\newblock \doi{10.1080/03610926.2025.2486538}.

\bibitem[{R Core Team}(2026)]{R_manual}
{R Core Team}.
\newblock \emph{R: A Language and Environment for Statistical Computing}.
\newblock R Foundation for Statistical Computing, Vienna, Austria, 2026.
\newblock URL \url{https://www.R-project.org/}.

\bibitem[Rajh-Weber et~al.(2025)Rajh-Weber, Huber, and
  Arendasy]{Rajh+et-al_2025}
H.~Rajh-Weber, S.~E. Huber, and M.~Arendasy.
\newblock A practice-oriented guide to statistical inference in linear modeling
  for non-normal or heteroskedastic error distributions.
\newblock \emph{Behavior Research Methods}, 57\penalty0 (12):\penalty0 338,
  2025.
\newblock \doi{10.3758/s13428-025-02801-4}.

\bibitem[Vasconcellos and Cribari-Neto(2026)]{Vasconcellos+Cribari_2026}
K.~L.~P. Vasconcellos and F.~Cribari-Neto.
\newblock A general approach to bias reduction and positive definiteness in
  covariance estimation.
\newblock \emph{Journal of Statistical Computation and Simulation}, 2026.
\newblock \doi{10.1080/00949655.2026.2694624}.
\newblock forthcoming.

\bibitem[Wachten and Strahler(2026)]{Wachten+Strahler_2026}
H.~Wachten and J.~Strahler.
\newblock Orthorexia nervosa and exercise addiction: distinct entities beyond
  restrictive and muscularity-oriented disordered eating behaviours?
\newblock \emph{Journal of Eating Disorders}, 14\penalty0 (1):\penalty0 34,
  2026.
\newblock \doi{10.1186/s40337-026-01535-8}.

\bibitem[White(1980)]{White_1980}
H.~White.
\newblock A heteroskedasticity-consistent covariance matrix estimator and a
  direct test for heteroskedasticity.
\newblock \emph{Econometrica}, 48\penalty0 (4):\penalty0 817--838, 1980.
\newblock \doi{10.2307/1912934}.

\end{thebibliography}

\end{document}